\documentclass[preprint]{elsarticle}
\journal{Nuclear Physics A}
\usepackage[latin9]{inputenc}
\usepackage[T1]{fontenc}
\usepackage[english]{babel}
\usepackage{amsmath}
\usepackage[inner=2.5cm,outer=2.5cm,top=0.5cm,bottom=1.6cm,includeheadfoot]{geometry}
\usepackage{graphicx}
\usepackage{subfigure}
\usepackage{cancel}
\usepackage{ulem}
\usepackage{amssymb}
\usepackage{epsfig}
\usepackage{appendix}
\usepackage{graphicx}
\usepackage{bm}
\usepackage{dcolumn}
\usepackage{multirow}
\usepackage{float}
\usepackage{listings}
\usepackage{color}
\usepackage{todonotes}
\usepackage{pdflscape}
\usepackage{appendix}

\include{srctex}

\newcommand{\vp}{\vec{p}\,}
\newcommand{\vl}{\vec{l}\,}
\newcommand{\hp}{\hat{p}\,}
\newcommand{\vq}{\vec{q}\,}

\newcommand{\vr}{\vec{r}}
\newcommand{\hr}{\hat{r}}
\newcommand{\vsu}{\vec{\sigma}_1}
\newcommand{\vsd}{\vec{\sigma}_2}
\newcommand{\vtu}{\vec{\tau}_1}
\newcommand{\vtd}{\vec{\tau}_2}

\newcommand{\ignore}[1]{}

\addto\captionsenglish{}

\begin{document}

\begin{frontmatter}

\title{Next-to-leading order effective field theory $\Lambda N\to NN$
potential in coordinate space}

\author[ujf]{A. P\'erez-Obiol}
\ead{perez-obiol@ujf.cas.cz}
\author[usal]{D. R. Entem}
\author[ubicc]{B. Juli\'a-D\'{\i}az}
\author[ubicc]{A. Parre\~no}

\address[ujf]{Nuclear Physics Institute, Czech Academy of Sciences,
250 68 \v{R}e\v{z}, Czech Republic}

\address[usal]{Grupo de F\'isica Nuclear and IUFFyM,
Universidad de Salamanca, E37008 Salamanca, Spain}

\address[ubicc]{Dept. d'Estructura i Constituents   de la Mat\`eria and  Institut de Ci\`encies del Cosmos (ICC),
Universitat de Barcelona, Mart\'{\i} i Franqu\`es 1, E08028-Spain}

\date{\today}

\begin{abstract}

The potential in coordinate space for the $\Lambda N\to NN$ weak transition,
which drives the weak decay of most hypernuclei,
is derived within the effective field theory formalism
up to next-to-leading order.
This coordinate space potential allows us 
to discuss how the different contributions to the potential add up at the different scales.
Explicit expressions are given for each two-pion-exchange diagram
contributing to the interaction.
The potential is also reorganized into spin and isospin operators,
and the coefficient for each operator is given in analytical form
and represented in coordinate space. The relevance of explicitly 
including the mass differences among the baryons appearing in the 
two-pion-exchange diagrams is also discussed in detail. 
\end{abstract}


\end{frontmatter}

\section{Introduction}

The $\Lambda N\to NN$ interaction is the main mechanism driving
the weak decay of heavy enough hypernuclei, see Refs.~\cite{botta2012,alberico02} 
for recent reviews. In this transition, each of the two final nucleons 
obtains a kinetic energy of about $\frac{1}{2}(M_\Lambda-M_N)= 88$ MeV, 
which allows them to escape or break any hypernucleus. This is in 
contrast with the mesonic decay, $\Lambda \to N\pi$, which is responsible 
for the weak decay of the lightest hypernuclei and of the $\Lambda$ in free 
space. In this latter case the final nucleon and pion share a kinetic 
energy of only $\sim M_\Lambda-M_N-m_\pi = 39$ MeV. For hypernuclei larger than 
$A=5$ this energy may not be enough for the nucleon to overcome the 
Pauli-blocking of the nuclear medium, and the one-nucleon induced decay mode
dominates the decay process. The two-nucleon induced mechanism, $\Lambda NN\to NNN$, 
also contributes to the weak decay, but it only represents about
20$\%$ of the non-mesonic weak decay amplitude~\cite{agnello}.

In hypernuclear experiments it is possible to measure the lifetime of 
a hypernucleus before it decays, as well as the angular and 
energetic distributions of the protons and neutrons emerging from the decay.
From these quantities one can obtain information about three independent observables that 
constrain the two-body $\Lambda N\to NN$ interaction: the total non-mesonic 
decay rate, $\Gamma_{\Lambda N\to NN}$, the neutron- or proton-induced decay 
rates, $\Gamma_{\Lambda n\to nn}$ or $\Gamma_{\Lambda p\to np}$, and the asymmetry 
between the intensities of protons going up and down the polarization 
axis of the hypernucleus. This last measurement is related to the interference 
between the parity-conserving (PC) and parity-violating (PV) parts of the
$\Lambda N\to NN$ amplitude. Using experimental data on the weak 
decay for both $s-$ and $p-$shell hypernuclei one is able to extract
at most six independent observables.

The weak $|\Delta S|=1$ $\Lambda N$ interaction may also be constrained,
in principle, by its experimental study in free space.
On the one hand through direct $\Lambda p$ scattering, and on the other 
through the $\Lambda$ production reaction $np\to \Lambda p$. However both possibilities 
present great experimental difficulties, the former one due to the small 
lifetime of the $\Lambda$ ($\tau_\Lambda=2.63\cdot 10^{-10}$ s) and the 
corresponding difficulty to produce stable $\Lambda$ beams, and the latter 
due to the very small cross section for the $np\to \Lambda p$ transition 
($10^{-12}$ mb)~\cite{Haidenbauer1995,Parreno1998,Inoue2001}. Currently we 
must rely on hypernuclear decay data to relate the theoretical description 
of this two-body interaction with the experiment. The experimental status 
and perspectives on non-mesonic weak decay has been recently reviewed 
in Ref.~\cite{botta15}.

Theoretical studies of the non-mesonic weak decay amplitude were first based on
one-meson-exchange (OME) models (see for 
example Refs.~\cite{mckellar84,nardulli88,dubach96,PRB97}). 
These models describe the long 
range part of the interaction through the exchange of one pion, and 
the shorter ranges through the exchange of heavier mesons, the
$\eta$, $\rho$, $\omega$, $K$ 
and $K^*$, which allow to mediate the strangeness exchange transition 
through weak vertices like $NNK$ or $\Lambda N\eta$. With the advent 
of the more systematic effective field theory (EFT) formalism, and in 
particular with its successful description of the NN strong 
interaction~\cite{entem,epelbaum}, first steps were done in applying 
EFT also to the description of the non-mesonic weak decay. The EFT for the 
$\Lambda N\to NN$ potential was first studied at leading order (LO) 
in Refs.~\cite{Jun,PBH05,axel1}. In Ref.~\cite{axel2} the EFT was 
further developed up to next-to-leading order (NLO), including all 
the possible two-pion-exchange (TPE) diagrams contributing to the transition. 
The potential was calculated in momentum space, and expressions in 
terms of master integrals were given separately for each diagram. 

In the present work we provide all the needed expressions in coordinate 
space. The different contributions to the transition potential are 
written in terms of 20 operational structures. Notably, a simplified 
version of the transition potential, obtained neglecting the baryonic 
mass differences of virtual baryons, provides a compelling description 
of the full potential. These should be readily useful for ab-initio 
few-body computations of the weak decay of hypernuclei ~\cite{hiyama09,nogga13,roth14,lonardoni14}. 

The manuscript is organized in the following way. In the beginning of 
Sect.~\ref{sec2} we review the EFT formalism used to calculate the 
non-mesonic weak transition. The EFT potentials up to NLO are derived 
in coordinate space and presented in terms of spin and isospin operators 
instead of diagrams. This allows us to plot for each operator the LO 
and the NLO potentials, and thus evaluate the magnitude of the 
two-pion exchanges in the $\Lambda N\to NN$ amplitude. The comparison is 
provided for each operational structure appearing in the transition potential. 
For the NLO, 
we derive approximate potentials neglecting the mass difference among 
the virtual baryons and compare them with the exact ones. In Sect.~\ref{sec3} 
we discuss the properties of the obtained potentials, including the comparison 
between the approximate and exact NLO potentials. A brief summary and 
conclusions are provided in Sect.~\ref{sec4}. Details of the calculation 
and the expressions for the potentials in coordinate space are provided in 
the Appendices.

\section{$\Lambda N\to NN$ EFT up to NLO}
\label{sec2}

The EFT potential for the $\Lambda N \to NN$ transition is built as an 
expansion on a parameter $\frac{q}{M}$, $q$ and $M$ representing the 
low and high energy scales that characterize the interaction.
These energy scales are determined by the typical momenta and masses
involved in the weak decay process. Since the reaction is exothermic 
($M_\Lambda>M_N$), the momenta of the emerging nucleons are larger
than the ones for the initial $\Lambda$ and nucleon. The momenta of the $\Lambda$ and the nucleon in the 
center of mass are labelled as $\vec{p}$ and $-\vec{p}$, and the momenta 
of the final two nucleons as $\vec{p}\;'$ and $-\vec{p}\;'$. The 
low-energy scale in the expansion parameter depends thus on two 
momenta, the initial momentum $\vec{p}$, of roughly $200$ MeV,
and the transferred momentum $\vec{q}\equiv \vec{p}\;'-\vec{p}$,
of the order of $400$ MeV.

Besides the $\Lambda$ and the nucleon we include the pion and the kaon
as virtual mesons to be exchanged among the baryons and the $\Sigma$ 
as an intermediate baryonic state. Therefore, the characteristic 
low-energy scales are also the pion and kaon masses and the mass 
differences between the $\Lambda$, the $\Sigma$ and the nucleon.
For the high-energy scale we take the chiral symmetry breaking scale, 
which is of the order of the mass of the baryons appearing
in the interaction, around $1000$ MeV.

Once the expansion parameter and the degrees of freedom are defined
we use Weinberg power-counting~\cite{W9091} to organize the potential
into the different orders.
At LO the potential includes the exchanges of the lightest 
pseudoscalar mesons and the non-derivative operators representing
contact interactions.
The $\eta$-exchange is not included due to its small coupling,
thus the one-meson-exchanges consist of the one-pion- and one-kaon-exchanges (OPE and OKE).
The NLO includes all the 
possible ways in which the $\Lambda$ and the nucleon can exchange two 
pions and the contact interactions containing one or two powers of 
momenta. The kaon-pion and the two-kaon exchange are
 not included in the NLO, first because 
their contribution is expected to be much smaller due to the larger mass 
of the kaon, and also to avoid adding further unknown (and therefore, 
model dependent) couplings to the theory.

\begin{figure}[tbh]
\centering
\begin{tabular}{cccc}
&
\includegraphics[clip=true,width=0.19\columnwidth]{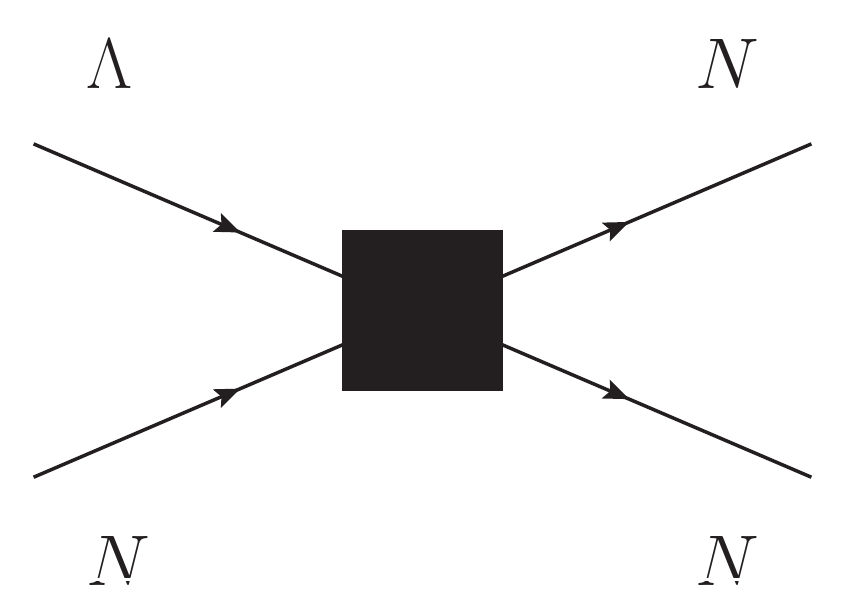}
&
\includegraphics[clip=true,width=0.20\columnwidth]{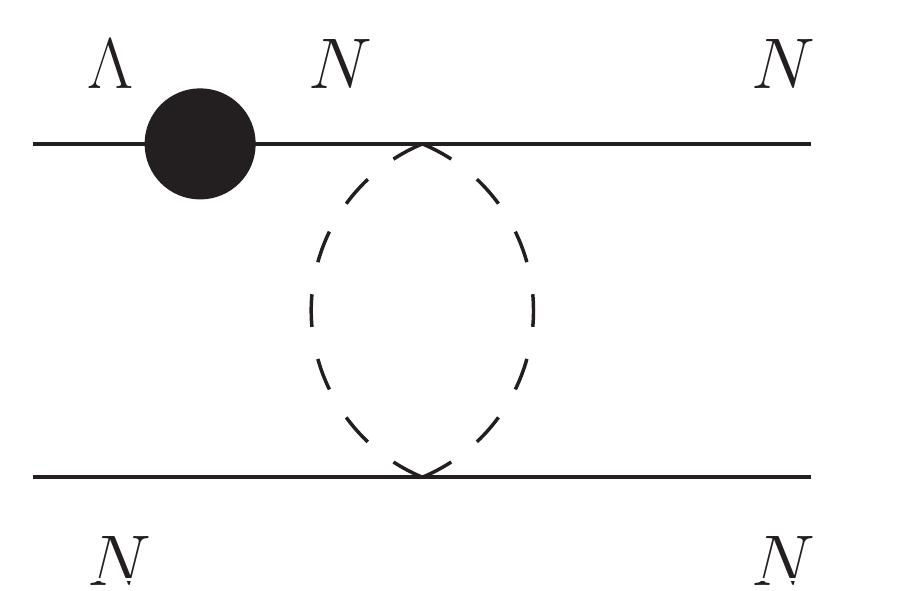}
&
\\
&  & (a) &
\\\\
\includegraphics[clip=true,width=0.20\columnwidth]{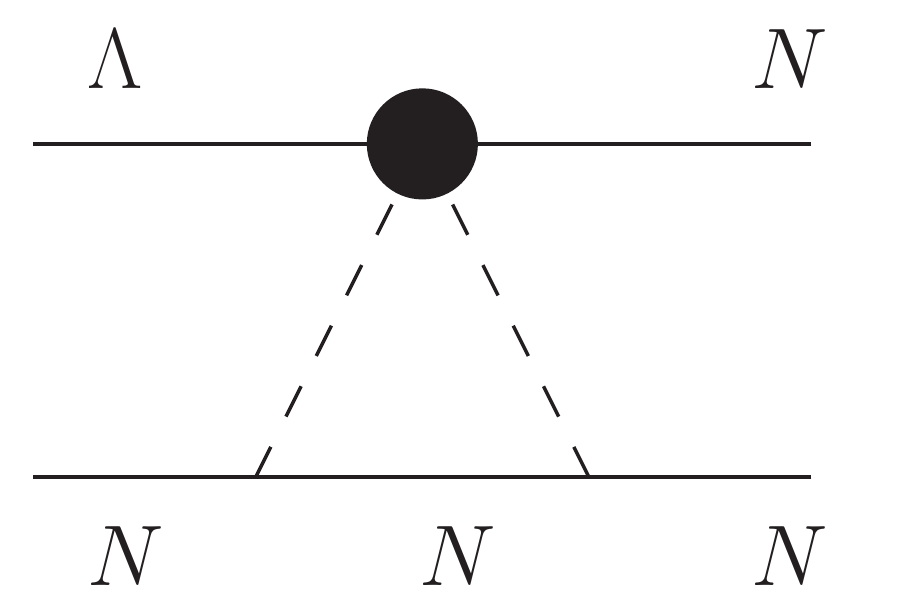}
&
\includegraphics[clip=true,width=0.20\columnwidth]{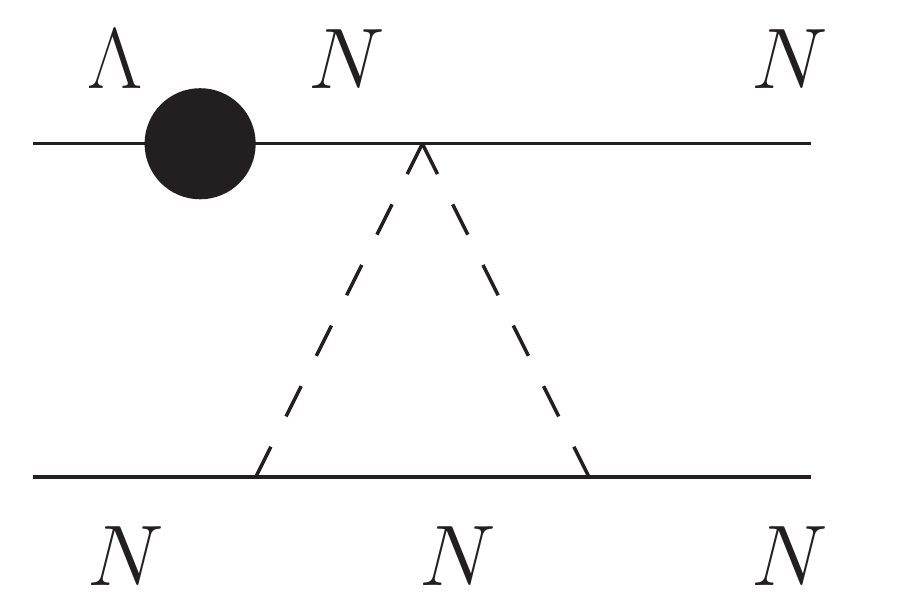}
&
\includegraphics[clip=true,width=0.20\columnwidth]{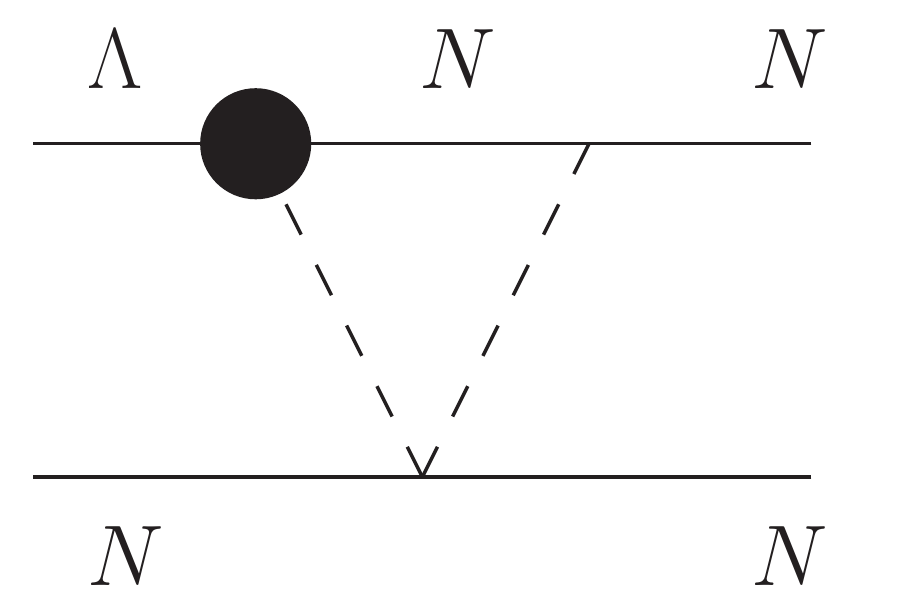}
&
\includegraphics[clip=true,width=0.20\columnwidth]{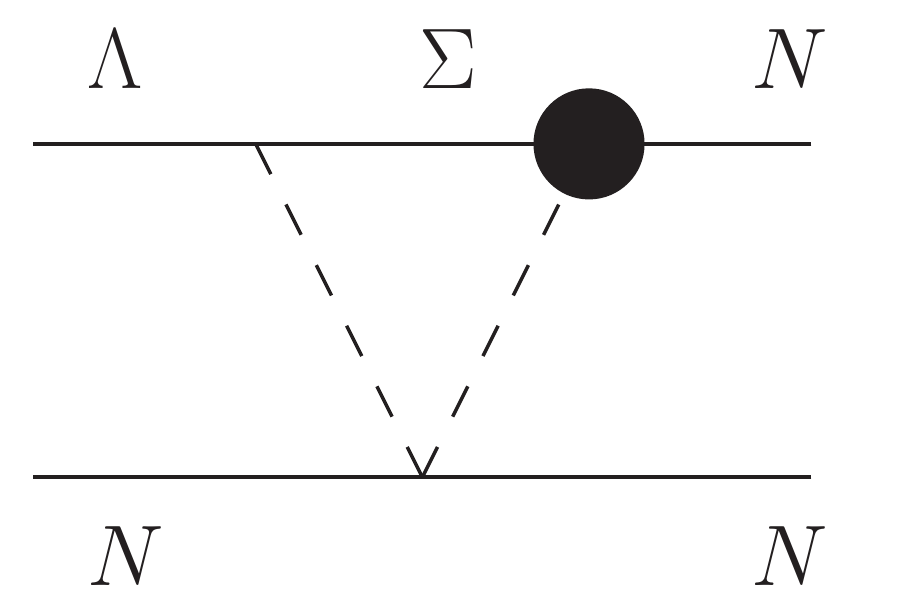}
\\(b)&(c)&(d)&(e)
\\\\
\includegraphics[clip=true,width=0.20\columnwidth]{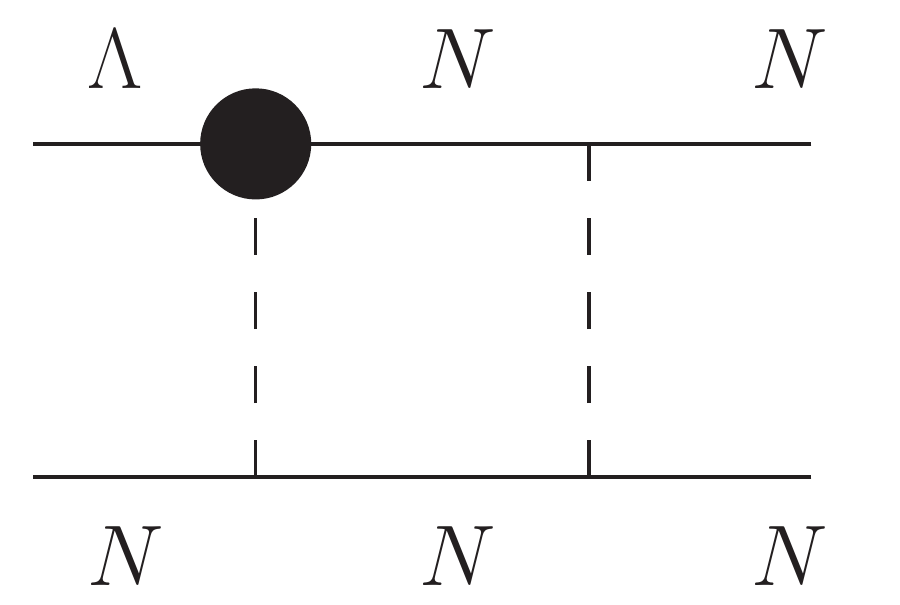}
&
\includegraphics[clip=true,width=0.20\columnwidth]{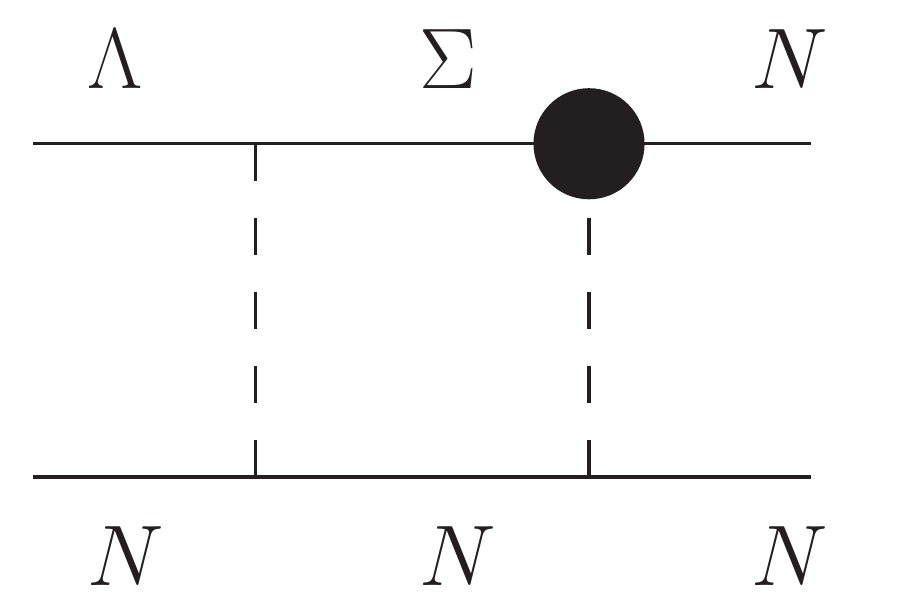}
&
\includegraphics[clip=true,width=0.20\columnwidth]{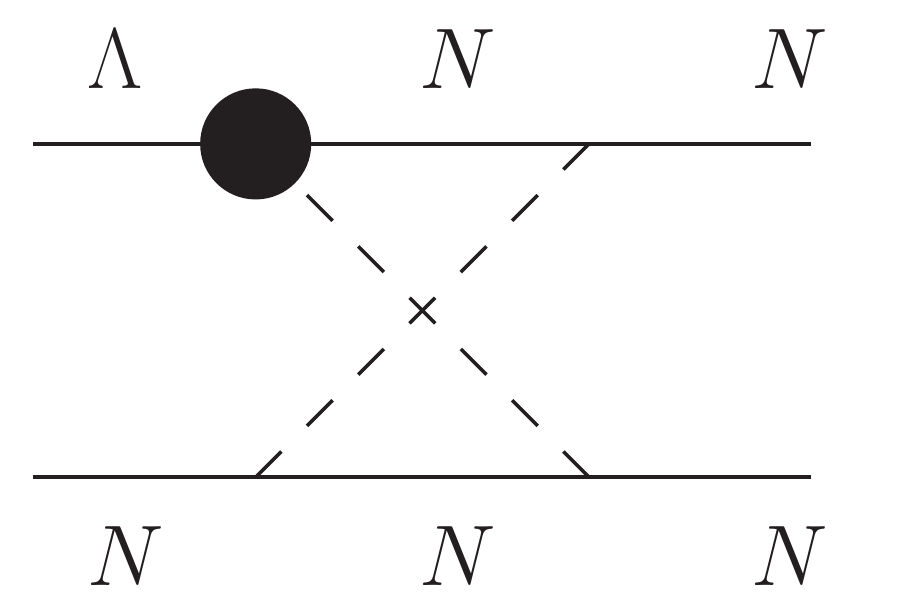}
&
\includegraphics[clip=true,width=0.20\columnwidth]{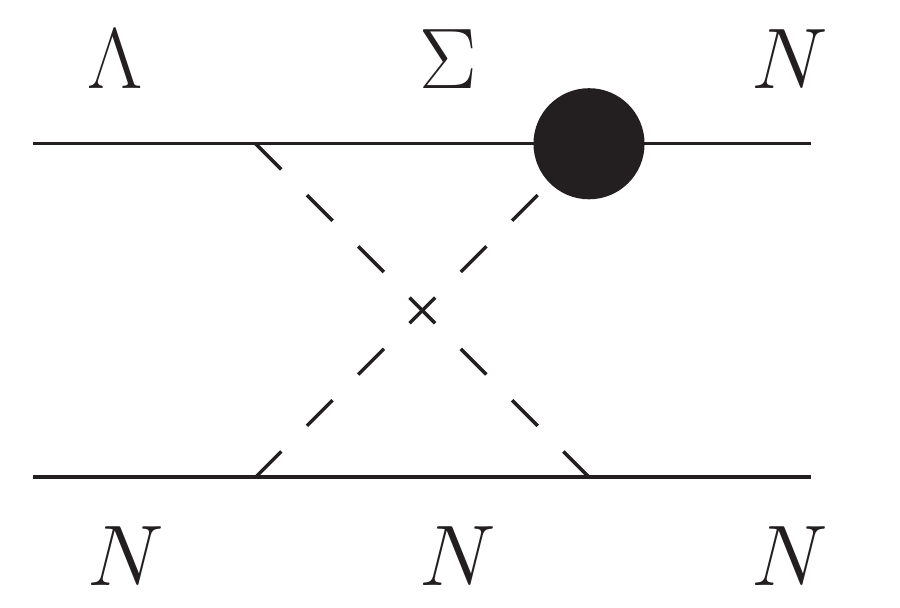}
\\(f)&(g)&(h)&(i)
\end{tabular}
\caption{Contact interactions and two-pion exchanges contributing
at next-to-leading order. The solid square represents operators at
order ${\cal O}(q)$ and ${\cal O}(q^2)$ and the solid vertex
represents the weak vertex.
\label{fig:nlodiagrams}}
\end{figure}

The Feynman diagrams corresponding to the LO and NLO are shown respectively
in Figs.~\ref{fig:lodiagrams} and \ref{fig:nlodiagrams}.
\begin{figure}[htb]
\centering
\begin{tabular}{ccc}
\includegraphics[clip=true,width=0.18\columnwidth]{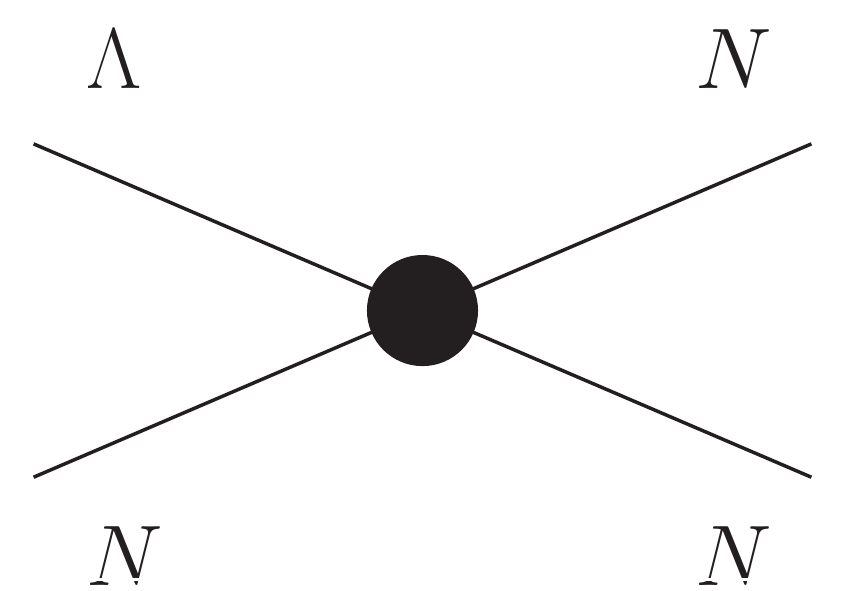}
&
\includegraphics[clip=true,width=0.18\columnwidth]{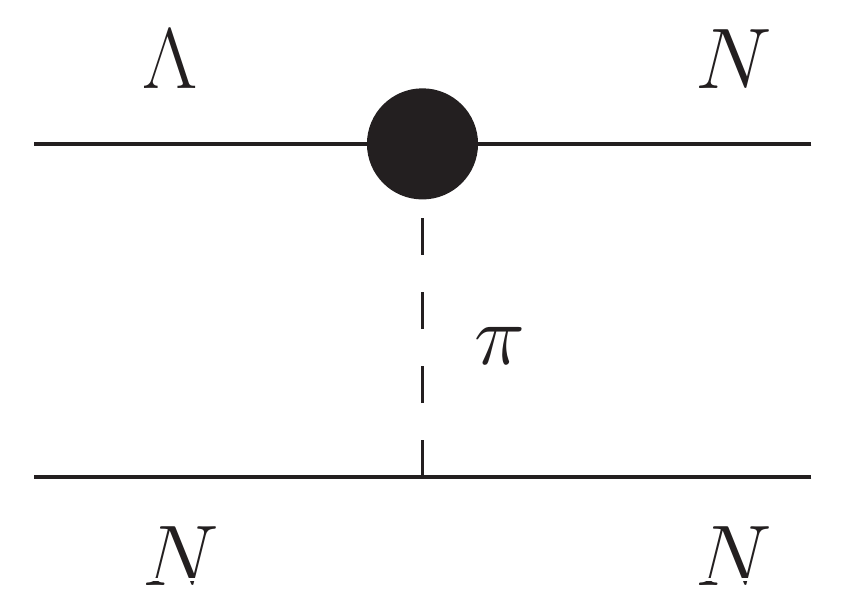}
&
\includegraphics[clip=true,width=0.18\columnwidth]{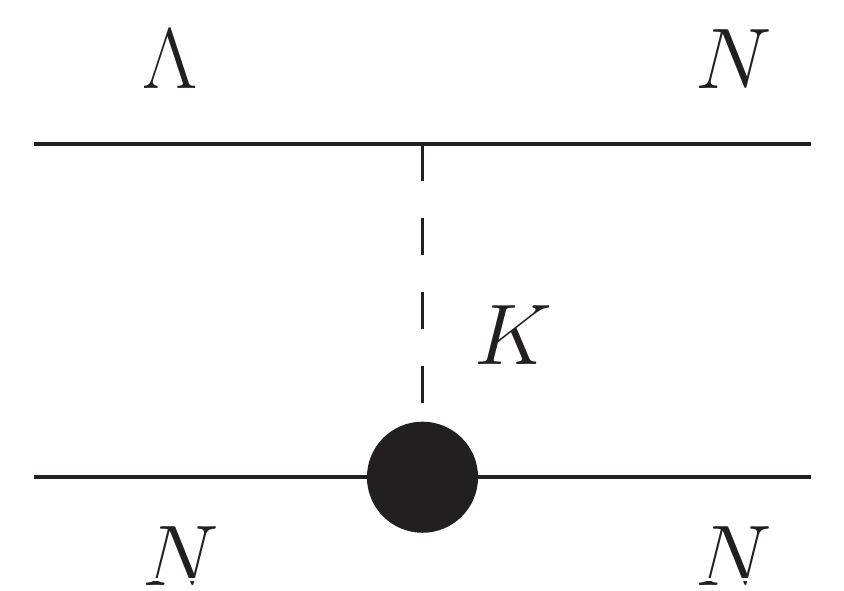}
\end{tabular}
\caption{Leading-order Feynman diagrams contributing to the 
$\Lambda N\to NN$ process. They correspond, in order, to the 
contact interactions, the one-pion exchange and the one-kaon 
exchange. The solid dot represents the weak vertex.\label{fig:lodiagrams}}
\label{fig:lodiagrams}
\end{figure}

The expressions for the potentials in momentum space for the 
$\Lambda N\to NN$ transition as well as details of their derivations 
can be found in Ref.~\cite{axel2}. They are given for each
possible diagram contributing to the transition and explicitly considering 
the mass differences between the $\Lambda$, the $\Sigma$ and the nucleon.
In this section we take these potentials and the mathematical formalism 
used in their derivation to calculate them in coordinate space.

\subsection{Leading order contributions}

At leading order there are only two operational structures contributing
to the contact part of the interaction, $\hat{1}$ and $\vsu\cdot\vsd$, and they 
are the same in momentum and coordinate space. $\vsu$ and $\vsd$ are the 
Pauli matrices connecting the spins of the upper and bottom baryons of 
the contact diagram in Fig.~\ref{fig:lodiagrams}. The one-pion and one-kaon 
exchanges in momentum space consist of a non-relativistic propagator, 
$\frac{1}{m_\alpha^2+\vec{q}\,^2}$, which depends on the mass and momentum 
of the meson, $m_\alpha$ and $\vec{q}$, and the parity-violating and tensor structures
$\vec{\sigma}_2\cdot\vec{q}$ and 
$(\vec{\sigma}_1\cdot\vec{q}\,)(\vec{\sigma}_1\cdot\vec{q}\,)$~\cite{axel1}.
They are straightforward to calculate in coordinate space by making the replacement 
$\vec{q}\to -i\vec{\nabla}$ in the Fourier transform.
For example, for a PV term $\vec{\sigma}\cdot\vq$ accompanying a
general function $f(q)$,
\begin{align}
\label{eq:ftpv}
\int \frac{d^3q}{(2\pi)^3}e^{i\vec{q}\cdot\vec{r}}f(q)\vec{\sigma}\cdot\vec{q}
=-i\vec{\sigma}\cdot\vec{\nabla}\int \frac{d^3q}{(2\pi)^3}e^{i\vec{q}\cdot\vec{r}}f(q).
\end{align}
The remaining integrand $f(q)$ only contains the propagator (multiplied
by couplings and masses), and thus one only needs to apply $\vec{\sigma}_2\cdot\vec{\nabla}$
and $(\vec{\sigma}_1\cdot\vec{\nabla})(\vec{\sigma}_2\cdot\vec{\nabla})$
to its Fourier transform, $\frac{e^{-m_\alpha r}}{4\pi r}$. We express both the 
OPE and the OKE potentials as
\begin{align}
\label{eq:omepot}
V_\alpha(r)=&
G_Fm_\pi^2\frac{e^{-m_{\alpha}r}}{12\pi r^3}
\left[
\hat{A}_\alpha 3r(1+m_{\alpha}r)\,i\vec{\sigma}_\alpha\cdot\hr
+\hat{B}_\alpha m_\alpha^2r^2\, \vec{\sigma}_1\cdot\vec{\sigma}_2
+\hat{B}_\alpha\left(3+3m_{\alpha}r+m_{\alpha}^2r^2\right)\hat{S}_{12}(\hat{r})
\right],
\end{align}
where the subindex $\alpha$ distinguishes between the pion ($\alpha=\pi$)
and the kaon ($\alpha=K$) exchanges and their corresponding operators, 
masses, and couplings. $\hat{S}_{12}(\hat{r})\equiv 3\vsu\cdot\hat{r}\vsd\cdot\hat{r}-\vsu\cdot\vsd$
is the tensor operator, and $\vec{\sigma}_\pi=\vec{\sigma}_2$, 
$\vec{\sigma}_K=\vec{\sigma}_1$. $G_Fm_\pi^2=2.21\cdot 10^{-7}$ is the Fermi 
constant, and $m_\pi=138$ MeV and $m_K=495$ MeV are the masses of the pion 
and the kaon. The isospin operators and the strong and weak couplings are 
encapsulated in $\hat{A}_\alpha$ and $\hat{B}_\alpha$, which are defined
as~\cite{PRB97}
$\hat{A}_\pi=-\frac{g_{NN\pi}A_\pi}{2M_N}\,\vec{\tau}_1\cdot\vec{\tau}_2$,
$\hat{B}_\pi=-\frac{g_{NN\pi}B_\pi}{4M_N\overline{M}}\,\vec{\tau}_1\cdot\vec{\tau}_2$,
$\hat{A}_K=\frac{g_{\Lambda NK}A_\pi}{2\overline{M}}\,
(\frac12 C_K^{PV}+D_K^{PV}+\frac12C_K^{PV}\vec{\tau}_1\cdot\vec{\tau}_2)$, and
$\hat{B}_K=-\frac{g_{\Lambda NK}A_\pi}{4M_N\overline{M}}\,
(\frac12 C_K^{PC}+D_K^{PC}+\frac12C_K^{PC}\vec{\tau}_1\cdot\vec{\tau}_2)$.
$M_N=939$ MeV is the mass of the nucleon, $\overline{M}=\frac12(M_\Lambda+M_N)=1027$
MeV is the average mass of the nucleon and the $\Lambda$, 
and $\vec{\tau}$ are the isospin Pauli matrices.
We take the values for the strong couplings $g_{NN\pi}=13.16$ 
and $g_{\Lambda NK}=-13.61$ from the Nijmegen 97f model \cite{nij99}.
The weak pionic couplings $A_\pi=1.05$ and $B_\pi=-7.15$ are fixed by the mesonic decay of the $\Lambda$,
while the kaonic ones $C_K^{PC}=-18.9$, $D_K^{PC}=6.63$, $C_K^{PC}=0.76$
and $D_K^{PC}=2.09$ are derived using SU(3) symmetry.

We have followed this procedure also in the NLO so both orders are 
calculated in the same way. However one may also regularize the OME 
potentials through form factors and directly Fourier-transform the 
whole expressions. The results only differ in the shortest range
of the interaction. Also note that in these expressions we have 
neglected the temporal part of the relativistic transferred momentum 
in the mesonic propagators, i.e. $q_0=\frac12(M_\Lambda-M_N)$ in
$\frac{1}{\vec{q}\,^2+m_\alpha^2-q_0^2}$. In order to take it into account
one only needs to replace $m_\alpha\to \sqrt{m_\alpha^2-q_0^2}$ in the 
expressions above. In this case the potentials would have the same 
Yukawa form but would decay slower due to their smaller effective masses.

\begin{figure}[t]
\center
\includegraphics[clip=true,width=1\hsize]{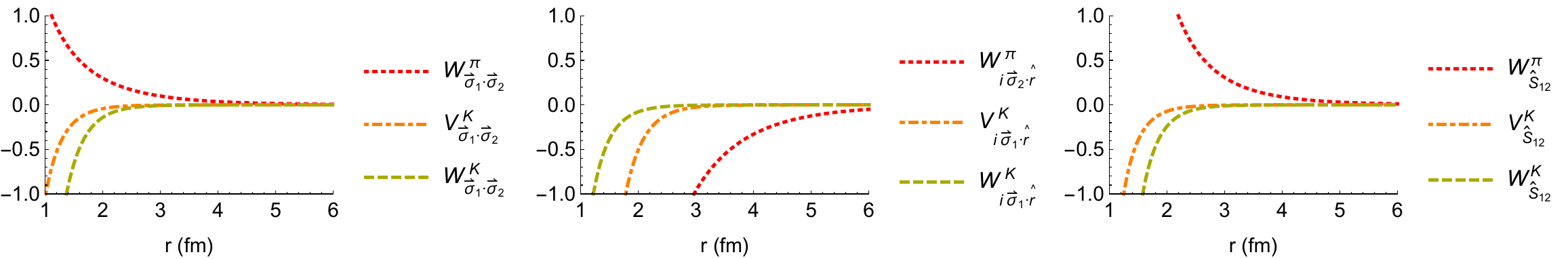}
\caption{Weak $\Lambda N\to NN$ potentials stemming from the one-pion and
one-kaon exchanges for the three spin operators to which they contribute.
The potentials are labelled $V$ and $W$ corresponding to the two isospin 
scalar operators $\hat{1}$ and $\vec{\tau}_1\cdot\vec{\tau}_2$. The potentials units 
are MeV, the momentum $|\vp|$ is fixed to $150$ MeV. }
\label{fig:omepot}
\end{figure}

In Fig.~\ref{fig:omepot} we plot the OPE and OKE potentials of 
Eq.~(\ref{eq:omepot}). The coefficients for the isospin operators $\hat{1}$ and 
$\vec{\tau}_1\cdot\vec{\tau}_2$ are labelled respectively $V$ and $W$.
For each of the three spin operational structures we plot the OPE, which 
only contains the isospin-isospin operator, together with the two isospin 
contributions from the OKE.

\subsection{Next-to-leading order contributions}

The contact part of the NLO potential contains all the possible operational 
structures that can be built with the momenta $\vec{p}$, $\vec{q}$ and the 
Pauli matrices $\vsu$ and $\vsd$. There are 18 in total ---6 at ${\cal O}(q)$ 
and 12 at ${\cal O}(q^2)$--- and they are shown in Table~\ref{tab:nlocontact}.
The corresponding operators in coordinate space are obtained by replacing 
$\vec{q}\to\vec{r}$ and $\vec{p}\to -i\vec{\nabla}$.
\begin{table}[t]
\centering
\begin{tabular}{c|c}
${\cal O}$($q$) &
$\vsu\cdot\vq$,
$\vsd\cdot\vq$,
$\vsu\cdot\vp$,
$\vsd\cdot\vp$,
$(\vsu\times\vsd)\cdot\vq$,
$(\vsu\times\vsd)\cdot\vp$,
\\\hline\\[-2.5ex]
${\cal O}$($q^2$) &
$\vq^2$,
$\vp^2$,
$\vq\cdot\vp$,
$(\vsu\cdot\vsd)\vq^2$,
$(\vsu\cdot\vsd)\vp^2$,
$(\vsu\cdot\vsd)\vq\cdot\vp$
\\&
$\vsu\cdot(\vq\times\vp)$,
$\vsd\cdot(\vq\times\vp)$,
$(\vsu\cdot\vp)(\vsd\cdot\vp)$,
\\&
$(\vsu\cdot\vq)(\vsd\cdot\vp)$,
$(\vsu\cdot\vp)(\vsd\cdot\vq)$,
$(\vsu\cdot\vq)(\vsd\cdot\vq)$.
\end{tabular}
\caption{
Contact NLO operators containing one and two powers of momenta $\vec{q}$, 
$\vec{p}$. $\vsu$ and $\vsd$ are the Pauli matrices.}
\label{tab:nlocontact}
\end{table}

The two-pion-exchange contribution to the NLO potential is much more involved
due to the various integrals appearing in the nine loop diagrams of 
Fig.~\ref{fig:nlodiagrams}. Terms corresponding to mass differences between 
the $\Lambda$, the $\Sigma$ and the nucleon also appear in the propagators 
of these NLO potentials, and neglecting these terms makes the expressions 
considerably simpler. However, it is not clear how these potentials change 
when these mass differences are neglected. In order to evaluate precisely 
this effect we derive in the next two subsections both the approximate and 
non-approximate potentials  and reorganize them in terms of operators.
Both cases contribute to 20 spin and isospin operational structures and are 
written in the following form:
\begin{align}
V(r)=&
\label{eq:nloops}
V_1+W_1(\vtu\cdot\vtd)
+V_{(\vsu\cdot\vsd)}(\vsu\cdot\vsd)+W_{(\vsu\cdot\vsd)}(\vsu\cdot\vsd)(\vtu\cdot\vtd)
\\&
+V_{(\hr\cdot\hp)}(\hr\cdot\hp)+W_{(\hr\cdot\hp)}(\hr\cdot\hp)(\vtu\cdot\vtd)
+V_{(\vsu\cdot\hr)}(\vsu\cdot\hr)+W_{(\vsu\cdot\hr)}(\vsu\cdot\hr)(\vtu\cdot\vtd)
\nonumber\\&
+V_{(\vsu\cdot\vsd)(\hr\cdot\hp)}(\vsu\cdot\vsd)(\hr\cdot\hp)
+W_{(\vsu\cdot\vsd)(\hr\cdot\hp)}(\vsu\cdot\vsd)(\hr\cdot\hp)(\vtu\cdot\vtd)
\nonumber\\&
+V_{(\vsu\cdot\hr)(\vsu\cdot\hp)}(\vsu\cdot\hr)(\vsu\cdot\hp)
+W_{(\vsu\cdot\hr)(\vsu\cdot\hp)}(\vsu\cdot\hr)(\vsu\cdot\hp)(\vtu\cdot\vtd)
\nonumber\\&
+V_{\vsu\cdot(\hr\times\hp)}\vsu\cdot(\hr\times\hp)
+W_{\vsu\cdot(\hr\times\hp)}\vsu\cdot(\hr\times\hp)(\vtu\cdot\vtd)
\nonumber\\&
+V_{\vsd\cdot(\hr\times\hp)}\vsd\cdot(\hr\times\hp)
+W_{\vsd\cdot(\hr\times\hp)}\vsd\cdot(\hr\times\hp)(\vtu\cdot\vtd)
\nonumber\\&
+V_{(\vsu\times\vsd)\cdot\hr}(\vsu\times\vsd)\cdot\hr
+W_{(\vsu\times\vsd)\cdot\hr}(\vsu\times\vsd)\cdot\hr(\vtu\cdot\vtd)
\nonumber\\&
+V_{\hat{S}_{12}}\hat{S}_{12}(\hat{r})
+W_{\hat{S}_{12}}\hat{S}_{12}(\hat{r})(\vtu\cdot\vtd).
\nonumber
\end{align}

\subsubsection{Potentials neglecting baryonic mass differences}

Neglecting $M_\Lambda-M_N$ and $M_\Sigma-M_\Lambda$ in the NLO potentials in 
momentum space of Ref.~\cite{axel2} we obtain their expressions in terms of
a set of integrals as shown in \ref{app:noq0potq}. These integrals 
are characteristic to the topologies and vertices of the diagrams contributing 
to the $\Lambda N\to NN$ transition. They depend on the number of propagators, 
the number of integrated momenta in the numerator, and the tensor structure 
---formed by Kronecker deltas $\delta_{ij}$ and transferred momenta 
$\vec{q}_i$--- to which they contribute. However, all of them can be 
related through Veltman-Passarino reductions~\cite{passarino} to the four 
following integrals,
\begin{align}
\label{eq:miqnoq0}
B(q)\equiv\,&
\frac1i\int \frac{d^4l}{(2\pi)^4}
\frac{1}{l^2-m^2+i\epsilon}
\frac{1}{(l+q)^2-m^2+i\epsilon}
=-\frac{1}{8\pi^2}L(q),
\\\nonumber
I(q)\equiv\,&
\frac1i\int \frac{d^4l}{(2\pi)^4}
\frac{1}{l^2-m^2+i\epsilon}
\frac{1}{(l+q)^2-m^2+i\epsilon}
\frac{1}{-l_0+i\epsilon}
=-\frac{1}{4\pi}A(q),
\\\nonumber
J(q)\equiv\,&
\frac1i\int \frac{d^4l}{(2\pi)^4}
\frac{1}{l^2-m^2+i\epsilon}
\frac{1}{(l+q)^2-m^2+i\epsilon}
\frac{1}{-l_0+i\epsilon}
\frac{1}{-l_0+i\epsilon}
=\frac{1}{2\pi^2}\frac{1}{4m^2+q^2}L(q),
\\\nonumber
K(q)\equiv\,&
\frac1i\int \frac{d^4l}{(2\pi)^4}
\frac{1}{l^2-m^2+i\epsilon}
\frac{1}{(l+q)^2-m^2+i\epsilon}
\frac{1}{-l_0-q_0'+i\epsilon}
\frac{1}{l_0+i\epsilon}
=-J(q)
+\frac{1}{2\pi q_0'}A(q),
\end{align}
where
\begin{align}
\label{eq:aqlq}
A(q)\equiv&\frac{1}{2|\vq|}\arctan\left(\frac{|\vq|}{2m}\right),
\\\nonumber
L(q)\equiv&\frac{\sqrt{4m^2+|\vq|^2}}{|\vq|}
\ln\left(\frac{\sqrt{4m^2+|\vq|^2}+|\vq|}{2m}\right).
\end{align}
These are the simplest integrals that appear, in order, in the ball, 
triangle, crossed box, and box diagrams. A constant term in $B$ has 
been obviated since its Fourier transform is ill defined and is already 
considered by the contact interactions. In these equations and in the 
rest of the manuscript we denote the mass of the pion $m$ and the 
relativistic transferred momentum $q$. The temporal part of $q$, $q_0$, 
depends on the baryonic mass differences and has been neglected.
$q_0'$ is of the same order than $q_0$ and in principle it appears in 
the baryonic propagators of the four integrals. It has been neglected 
in all of them except in the last one, $K$, in order to avoid the pinch 
singularity characteristic of the box diagrams. In the final result of 
$K$ $q_0'$ is also neglected except in the divergent term $\frac{1}{q_0'}$.

\begin{figure}[t] 
\center
\includegraphics[clip=true,width=1\columnwidth]{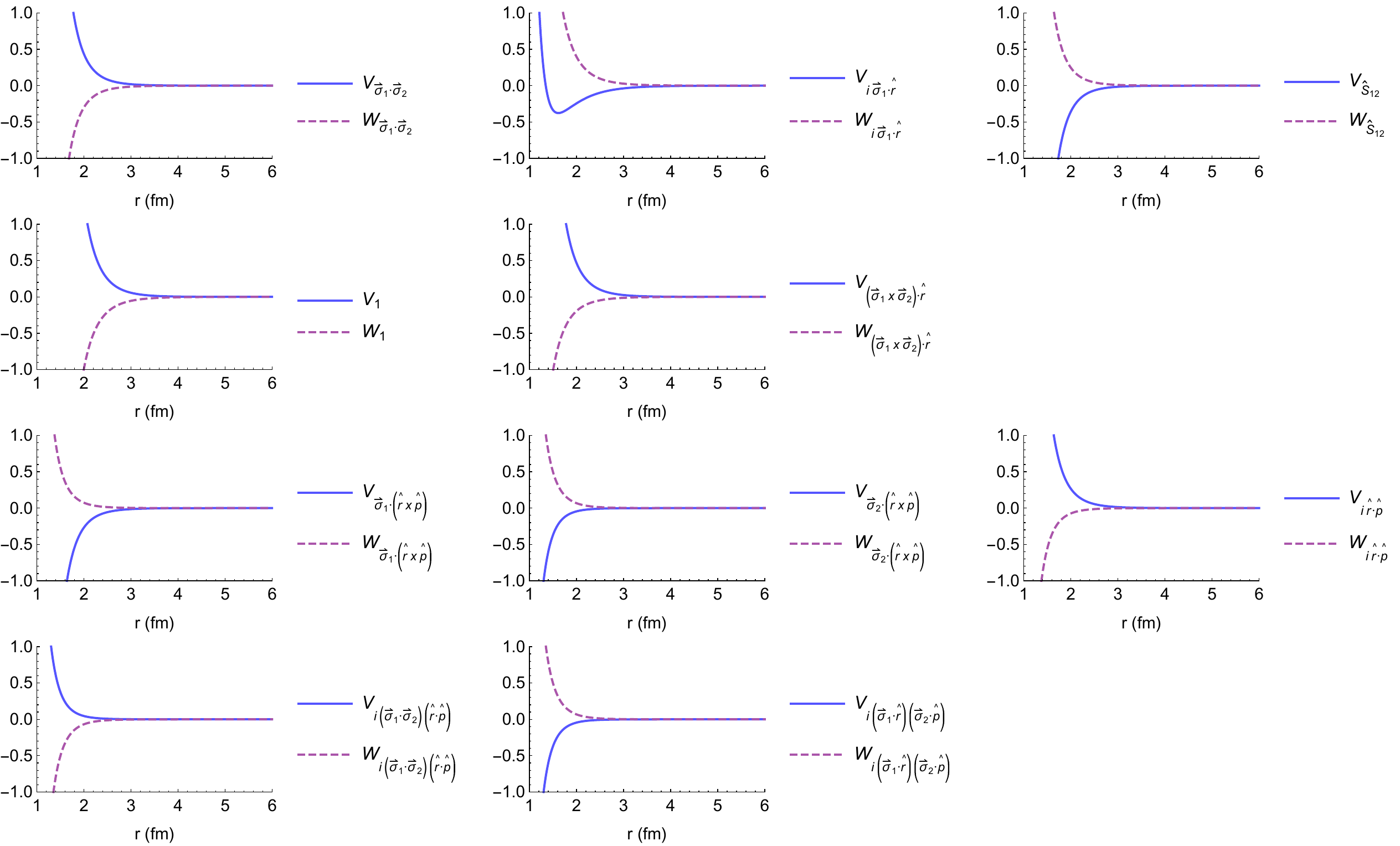}
\caption{Weak $\Lambda N\to NN$ potentials arising from the two-pion exchanges.
The units of the potentials are MeV. $|\vp|$ is fixed to $150$ MeV. 
}
\label{fig:tpepotnoq0}
\end{figure}

The potentials thus depend only on $A(q)\vq^{2n}$, $L(q)\vq^{2n}$ and 
$\frac{L(q)}{4m^2+q^2}\vq^{2n}$, where $n=0,1,2$, and the same terms 
but accompanied by spin structures (see \ref{app:noq0potr}).
To calculate the potentials in coordinate space we only need to Fourier 
transform this set of functions. All these Fourier transforms can be 
written in terms of exponentials and Bessel functions of order zero and one 
that depend on $2mr$ ($K_0(2mr)$ and $K_1(2mr)$).
For example, for the simplest cases we have:
\begin{align}
\label{eq:exftaqlq}
\mathcal{F}\left[A(q)\right]&=
\frac{e^{-2 m r}}{8 \pi r^2},
\\\nonumber
\mathcal{F}\left[L(q)\right]&=
-\frac{m}{2\pi r^2} K_1(2mr),
\\\nonumber
\mathcal{F}\left[\frac{L(q)}{4m^2+q^2}\right]&=
\frac{1}{4\pi r} K_0(2mr).
\end{align}
The explicit expressions for each operator coefficient of Eq.~(\ref{eq:nloops}) 
in terms of the Bessel functions and exponentials are shown in 
\ref{app:noq0potr}. We plot these coefficients in 
Fig.~\ref{fig:tpepotnoq0}. For each spin operator we have the two isospin scalar contributions,
$\hat{1}$ and $\vec{\tau}_1\cdot\vec{\tau}_2$, labelled as $V$ and $W$.

\subsubsection{Potentials considering baryonic mass differences}
\begin{figure}[t]
\center
\includegraphics[clip=true,width=1\columnwidth]{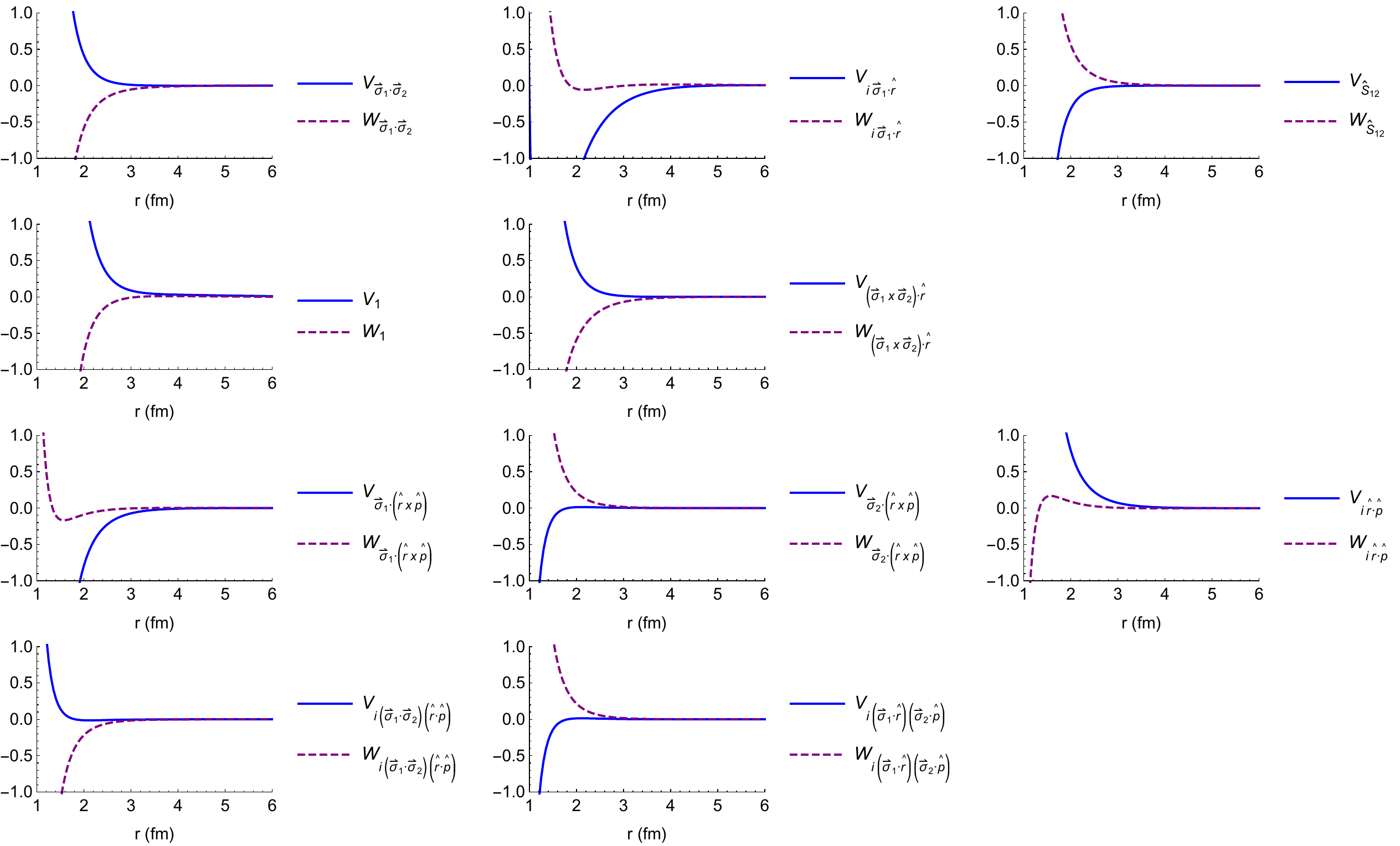}
\caption{Real part of the two-pion-exchange potentials for the 
$\Lambda N\to NN$ transition. The units of the potentials are MeV, $|\vp|$ 
is fixed to 150 MeV. }
\label{fig:tpepotq0r}
\end{figure}

\begin{figure}[t]
\center
\includegraphics[clip=true,width=1\columnwidth]{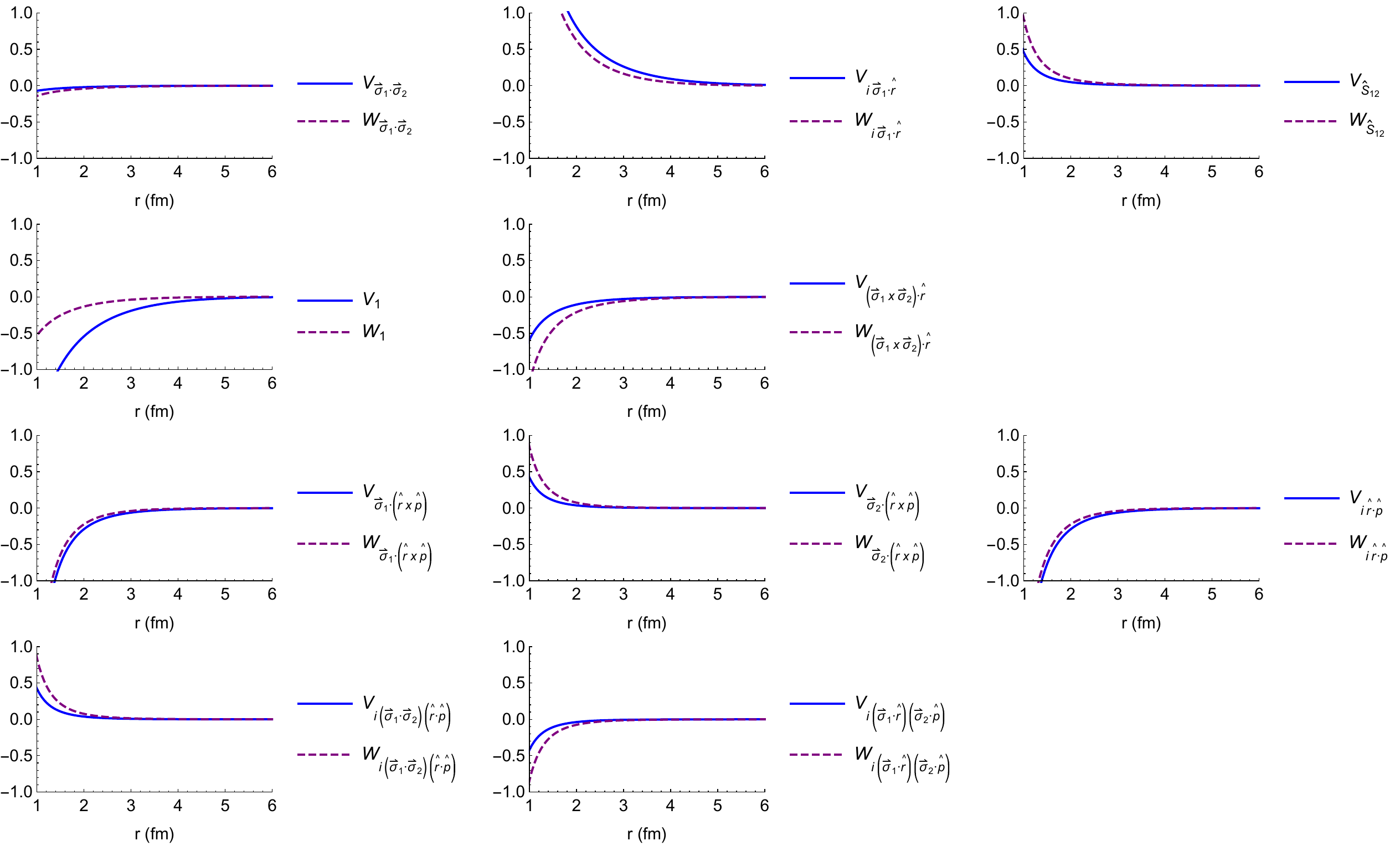}
\caption{Imaginary part of the two-pion-exchange potentials for the $\Lambda N\to NN$ 
transition. The units of the potentials are MeV, $|\vp|$ is fixed to 150 MeV.}
\label{fig:tpepotq0i}
\end{figure}

In the case in which the terms $q_0$ and $q_0'$ are explicitly considered it is easier
to calculate the necessary integrals directly in coordinate space.
These are labelled again according to the diagram in which they appear,
namely $B$, $I$, $J$ and $K$ for the ball, triangle, crossed-box, and box
diagrams. The relativistic subindices $\mu...\nu$ correspond
to the momenta appearing in the numerator:
\begin{align}
\label{eq:mirq0}
B_{\mu...\nu}(q)=&
\frac1i\int\frac{d^3q}{(2\pi)^3}e^{i\vec{q}\cdot\vec{r}}\int \frac{d^4l}{(2\pi)^4}
\frac{1}{l^2-m^2+i\epsilon}
\frac{1}{(l+q)^2-m^2+i\epsilon}l_\mu...l_\nu,
\\\nonumber
I_{\mu...\nu}(q)=&
\frac1i\int\frac{d^3q}{(2\pi)^3}e^{i\vec{q}\cdot\vec{r}}\int \frac{d^4l}{(2\pi)^4}
\frac{1}{l^2-m^2+i\epsilon}
\frac{1}{(l+q)^2-m^2+i\epsilon}
\frac{1}{-l_0-q_0'+i\epsilon}l_\mu...l_\nu,
\\\nonumber
J_{\mu...\nu}(q)=&
\frac1i\int\frac{d^3q}{(2\pi)^3}e^{i\vec{q}\cdot\vec{r}}
\int \frac{d^4l}{(2\pi)^4}
\frac{1}{l^2-m^2+i\epsilon}
\frac{1}{(l+q)^2-m^2+i\epsilon}
\frac{1}{-l_0-q_0'+i\epsilon}
\frac{1}{-l_0+i\epsilon}l_\mu...l_\nu,
\\\nonumber
K_{\mu...\nu}(q)=&
\frac1i\int\frac{d^3q}{(2\pi)^3}e^{i\vec{q}\cdot\vec{r}}
\int \frac{d^4l}{(2\pi)^4}
\frac{1}{l^2-m^2+i\epsilon}
\frac{1}{(l+q)^2-m^2+i\epsilon}
\frac{1}{-l_0-q_0'+i\epsilon}
\frac{1}{l_0+i\epsilon}l_\mu...l_\nu.
\end{align}
When the integrals also contain transferred momenta $\vec{q}_i$
in the numerators we can make the replacement $\vec{q}_i\to-i\vec{\nabla}_i$
and calculate them in terms of derivatives of the previous ones.
These integrals will contain tensor structures formed by
Kronecker deltas and vectors $\vec{r}_i$. The coefficients for
each structure are defined in \ref{app:q0mir}, and the
potentials for each diagram are shown in \ref{app:q0pot}.
We reorganize again these potentials in terms of the spin operators of
Eq.~(\ref{eq:nloops}) and show the corresponding expressions
also in \ref{app:q0pot}.
In contrast with the case in which $M_\Lambda-M_N$ is neglected,
now the potentials contain imaginary parts.
This is because the initial masses are larger than the final ones
and therefore the process is not unitary.
In figs.~\ref{fig:tpepotq0r} and~\ref{fig:tpepotq0i} we plot the real and 
imaginary parts of the NLO potentials for each operator, respectively.

\section{Coordinate space potentials}
\label{sec3}

In this section we first discuss how the LO and NLO potentials contribute 
to the different spin-isospin operational structures listed in 
Table~\ref{tab:nlocontact}. Secondly, we analyze the effect of neglecting 
the baryonic mass differences in the baryonic propagators. 

\subsection{Contributions from the OPE, OKE and TPE}

The LO potential is represented in Fig.~\ref{fig:omepot}
and the NLO one in Figs.~\ref{fig:tpepotnoq0},
\ref{fig:tpepotq0r}, and~\ref{fig:tpepotq0i}.
In the long range the $\Lambda N\to NN$ potential is
clearly dominated by the one-pion-exchange.
The one-kaon exchange is shorter ranged and
interferes destructively with the OPE in the spin-spin and tensor
potentials and constructively in the PV one.

At NLO the potential contains many more operational structures
than the LO OPE and OKE, which allows for many more possible spin
$\Lambda N\to NN$ transitions.
For each spin operator the two isospin scalar contributions
contribute with opposite signs as can be noticed in Fig.~\ref{fig:tpepotnoq0}.

\begin{figure}[t]
\center
\includegraphics[clip=true,width=1\columnwidth]{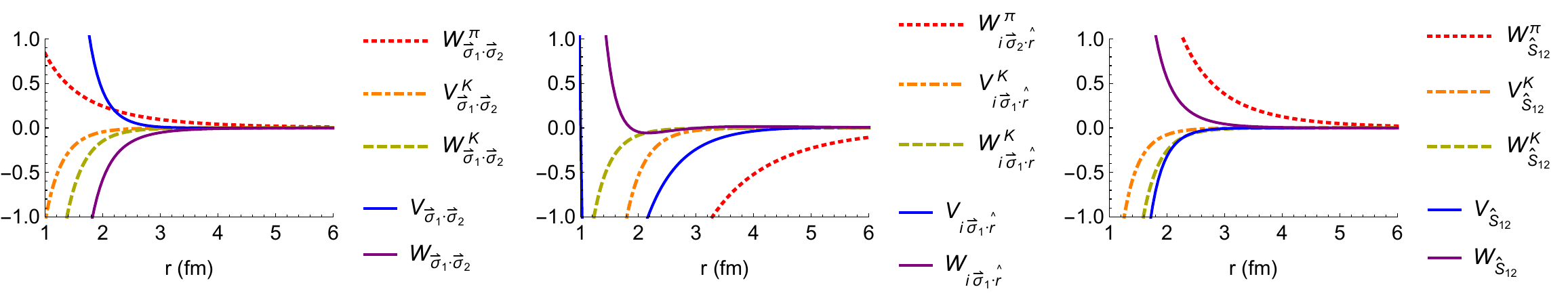}
\caption{One-meson and two-pion exchange potentials for the three 
LO spin operators. For each of these operators the OPE and OKE are 
labelled with the superindices $\pi$ and $K$, and the scalar and 
isospin-isospin two-pion exchanges with $V$ and $W$. The units of the 
potentials are MeV, $|\vp|$ is fixed to 150 MeV. }
\label{fig:opetpepot}
\end{figure}

In Fig.~\ref{fig:opetpepot} we plot together the LO and the NLO potentials
for each of the three LO spin operators.
The typical range of the two-pion exchanges is shorter than
the one-meson exchanges.
Above 2 fm the pion dominates the interaction,
while in shorter ranges the NLO potential may surpass it, as seen
for example in the spin-spin case.
The one-kaon exchange decays faster than the one-pion exchange
and is already dominated by the NLO in the medium range.
For each operator the two-pion exchanges 
represent a noticeable part of the $\Lambda N\to NN$ amplitude.
Moreover for certain transitions the 20 spin and isospin operators
may interfere constructively, rendering a greater importance
to the two-pion exchange than the one inferred from Fig.~\ref{fig:opetpepot}.

\subsection{NLO potentials in the $M_\Lambda=M_\Sigma=M_N$ limit}

In the SU(3) chiral effective Lagrangian the different octet masses appear
at second order due to the explicit symmetry breaking term involving 
different quark masses. One can consider then an average mass for the 
baryon $\overline{M}$ plus a term of order ${\cal O}(q^2)$. With this 
approximation the potentials for the $\Lambda N\to NN$ transition are 
purely real, since the initial and final masses are considered the same. 
Besides this feature, the explicit effect on the real part of the potentials
is difficult to predict due to the complexity of the calculation.

\begin{figure}[t]
\center
\includegraphics[clip=true,width=1\columnwidth]{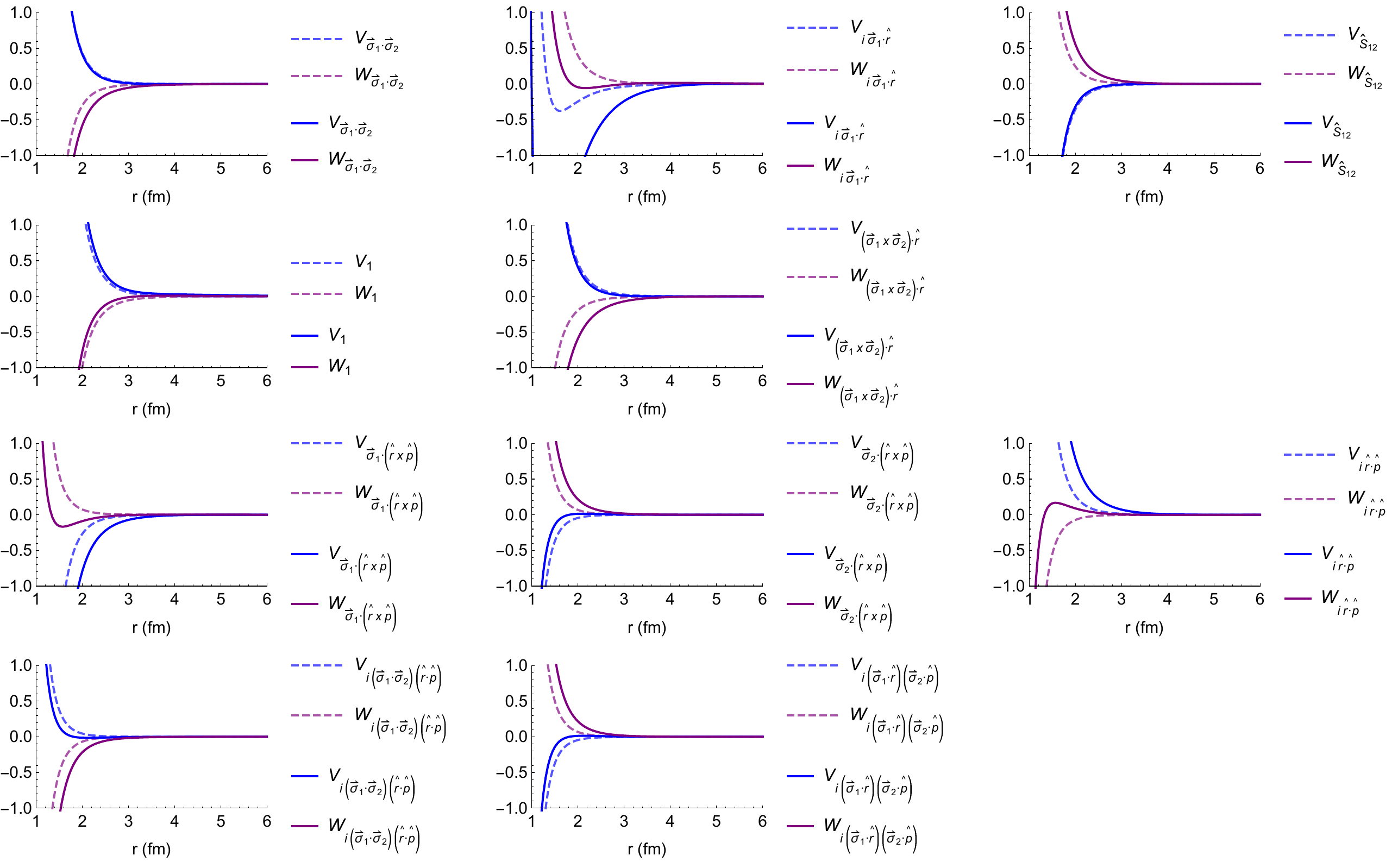}
\caption{Comparison between the real part of the potentials with
$q_0=q_0'=0$ (dashed lines) and with $q_0\neq0$, $q_0'\neq0$ (solid lines).
The potential units are MeV, $|\vp|$ is fixed to 150 MeV.}
\label{fig:tpeq0noq0}
\end{figure}

In Fig.~\ref{fig:tpeq0noq0} we have plotted together the real parts of 
the NLO potentials obtained with and without the approximation. 
Only in one of the twenty operational structures we see a relevant
qualitative change in the potentials. All of them maintain its 
attractive or repulsive character,
however the PV isospin scalar $\vsu\cdot\hr$ structure
becomes much more attractive when the different 
masses are considered.
Except for that amplitude, the changes in all other 
structures are also fairly small from a quantitative point of view. 
We do not find any common pattern
---some become slightly smaller and some slightly larger---. From a practical 
point of view, the NLO potentials obtained considering the terms 
$q_0$ and $q_0'$ are certainly more precise, but also much more complicated
and computationally time consuming to calculate. Thus, we expect that 
for any practical application to compute non-mesonic weak decay rates,
the approximate potentials should be accurate enough.

\subsection{Contact part of the interaction}

The EFT potential for the weak $|\Delta S|=1$ $\Lambda N$ interaction
includes de medium and long range OPE, OKE and TPE diagrams
as well as a series of contact interactions represented by the 
terms of Table~\ref{tab:nlocontact}. The coefficients of these 
terms, or the low energy constants (LECs), describe the short 
range physics and must be fixed by experiment. In Ref.~\cite{axel1} 
the LO LECs were constrained using recent
experimental data on non-mesonic weak decay of s- and p-shell hypernuclei.
It was found that different sets of values of the LO LECs
could fit the data with a reasonable $\chi^2$.
With the current status of experimental data for the non-mesonic
weak decay of hypernuclei it is not possible to fix
the 20 LECs characteristic of this interaction.
Even with more accurate data there will always remain the difficulty
to take into account the nuclear medium
when constraining all the possible two-body PC and PV transitions.
As already motivated in Ref.~\cite{axel2} experimental data
from the inverse reaction $np\to \Lambda p$ would be of great value to fix
the LO and NLO LECs. However, this study may guide more phenomenological 
approaches, giving more or less importance to the different possible 
transitions according to the two-pion exchanges, which are now explicitly known.

\section{Summary and conclusions}
\label{sec4}

In this manuscript we have calculated the potentials in coordinate 
space for the $\Lambda N\to NN$ interaction within the EFT formalism 
and up to NLO. For the contact part of the interaction we have 
constructed all the possible spin and isospin operational structures 
up to two powers of momenta. These consisted in the two LO structures 
$\hat{1}$ and $\vsu\cdot\vsd$ and the 18 NLO ones of Table~\ref{tab:nlocontact}. 
The LO one-pion and one-kaon exchanges contribute to only six of 
these operators, the spin-spin $(\vsu\cdot\vsd)$, the tensor $\hat{S}_{12}(\hat{r})$,
the parity-violating $\vsu\cdot\hat{r}$ and the same ones
multiplied by $\vec{\tau}_1\cdot\vec{\tau}_2$. These contributions have 
the form of Yukawa potentials and their derivatives and describe the 
longest range of the interaction.

To calculate the NLO potentials coming from the two-pion-exchange 
diagrams we have first approximated them by neglecting the mass 
differences among the baryons entering in the transition. In this 
case the calculation consists in Fourier transforming a basic set 
of functions and then reorganize the potential in terms of operational 
structures. For the more complicated case in which the 
different baryonic masses are explicitly considered we have calculated 
from the beginning the loop integrals in coordinate space. Both potentials 
are organized into the 20 different spin and isospin operators of 
Eq.~(\ref{eq:nloops}).

The two-pion exchange contributions have been shown to contribute to a 
large number of operational structures. Their effect should thus be 
sizeable and difficult to mimic by other means, e.g. the correlated 
part could be partly understood as a one-sigma exchange potential, but 
all other contributions will appear in many-different partial waves.
They should, eventually, help to understand more precise data on non-mesonic
weak decay, in particular from the experiments E18~\cite{jparce18}
(on $^{12}_\Lambda \text{C}$) and E22~\cite{jparce22} (on $A=4$ hypernuclei) in J-PARC.
From a practical 
point of view the potentials obtained neglecting the baryonic mass differences 
should be accurate and fast enough to be amenable for state-of-the-art 
few-body computations.

\section{Acknowledgments}

This work was partly supported by the GA\v{C}R grant P203/15/04301S (Czech Republic),
by the Ministerio de Econom\'ia y Competitividad under Contract Nos. FPA2013-47443-C2-2-P and FIS2014-54672-P (Spain),
by the Generalitat de Catalunya under grant 2014 SGR 401,
and by the Spanish Excellence Network on Hadronic Physics FIS2014-57026-REDT.
B.J-D. is supported by the Ramon y Cajal program.

\appendix

\section{Fourier transform of the potential for $q_0=q_0'=0$}
\label{app:noq0}

In this Appendix we explicitly show the details of the derivation
of the $\Lambda N\to NN$ potentials in coordinate space
when the baryonic mass differences are neglected ($q_0=q_0'=0$).
First we simplify the potentials of Ref.~\cite{axel2}
using that $q_0=q_0'=0$.
In this case the potentials only depend on a basis set of functions
and operators that can be solved analytically.
In the second section we Fourier transform these functions
and give the analytical expressions
of the potentials for each spin operator contributing to the transition.

\subsection{Potential in momentum space}
\label{app:noq0potq}
The simplest integrals appearing in the ball, triangle,
crossed box and box diagrams are labeled respectively as
$B$, $I$, $J$ and $K$ as shown in Eq.~(\ref{eq:miqnoq0}).
When these integrals have also momenta in the numerators 
they are labeled with the subindices of the momenta.
For example, when $J$ has $l_\mu$ in the numerator we label it
$J_\mu$, where $\mu$ is a relativistic subindex.
The results of these integrals are organized according
to the various structures formed by vectors $\vq$ and Kronecker deltas
to which they contribute.
The coefficients of these structures are defined in the same way for the
four types of integrals, therefore we only need to define them once.
Particularly, for the integrals appearing in the crossed-box
diagrams we have
\begin{align}
\label{eq:miqorg}
J_\mu\equiv\,&
\delta_{\mu0}J_{10}+\delta_{\mu
i}J_{11}\vq_i 
\,,\\\nonumber
J_{\mu\nu}\equiv\,&
\delta_{\mu0}\delta_{\nu0}J_{20}
+(\delta_{\mu0}\delta_{\nu i}
+\delta_{\mu i}\delta_{\nu 0})J_{21}\vq_i
+\delta_{\mu i}\delta_{\nu j}(J_{22}\delta_{ij}
+J_{23}\vq_i\vq_j)\,,
\\\nonumber
J_{\mu\nu\rho}\equiv\,&
\delta_{\mu0}\delta_{\nu0}\delta_{\rho0}J_{30}
+\delta\delta\delta_{\{\mu\nu\rho 00i\}}\vq_iJ_{31}
+\delta\delta\delta_{\{\mu\nu\rho 0ij\}}
(\delta_{ij}J_{32}+\vq_i\vq_jJ_{33})
\\&
+\delta_{\mu i}\delta_{\nu j}\delta_{\rho k}
(\delta\vq_{\{ijk\}}J_{34}+\vq_i\vq_j\vq_kJ_{35})\,,
\nonumber\\\nonumber
J_{\mu\nu\rho\sigma}\equiv\,&
\delta_{\mu0}\delta_{\nu0}\delta_{\rho0}\delta_{\sigma0}J_{40}
+\delta\delta\delta\delta_{\{\mu\nu\rho\sigma000i\}}\vq_iJ_{41}
+\delta\delta\delta\delta_{\{\mu\nu\rho\sigma00ij\}}
(\delta_{ij}J_{42}+\vq_i\vq_jJ_{43})
\\&
+\delta\delta\delta\delta_{\{\mu\nu\rho\sigma0ijk\}}
(\delta\vq_{\{ijk\}}J_{44}+\vq_i\vq_j\vq_kJ_{45})
\nonumber\\&
+\delta_{\mu i}\delta_{\nu j}\delta_{\rho k}\delta_{\sigma l}
(\delta\delta_{\{ijkl\}}J_{46}
+\delta\vq\vq_{\{ijkl\}}J_{47}
+\vq_i\vq_j\vq_k\vq_lJ_{48})\,, \nonumber
\end{align}
where
\begin{align}
\label{eq:deltaqstr}
\delta\vq_{\{ijk\}}=&\,
\delta_{ij}\vq_k
+\delta_{ik}\vq_j
+\delta_{jk}\vq_i\,,
\\\nonumber
\delta\vq\vq_{\{ijkl\}}=&\,
\delta_{ij}\vq_k\vq_l
+\delta_{ik}\vq_j\vq_l
+\delta_{il}\vq_j\vq_k 
+\delta_{jk}\vq_i\vq_l
+\delta_{jl}\vq_i\vq_k
+\delta_{kl}\vq_i\vq_j\,,
\\\nonumber
\delta\delta_{\{ijkl\}}=&\,
\delta_{ij}\delta_{kl}
+\delta_{ik}\delta_{jl}
+\delta_{il}\delta_{jk}\,.
\end{align}
The quantities, $\delta\delta\delta_{\{\mu\nu\rho00i\}}$,
$\delta\delta\delta_{\{\mu\nu\rho0ij\}}$, etc,
 indicate how many of the indices $\mu$,
$\nu$, $\rho$, and $\sigma$ must be temporal and how many spatial,
and they should not be
contracted with the indices $i$, $j$, and $k$ appearing in the rest of
the expressions.

The various coefficients in front of every tensor,
as $J_{10}$ or $J_{11}$, can be related
to the simpler integrals of Eq.~(\ref{eq:miqnoq0})
 using Veltman-Passarino tricks
---namely making contractions with $\vec{q}$ and adding and subtracting
terms in the numerators of the integrals---.
The necessary relations for the ball integrals are:
\begin{align}
\label{eq:miqrelb}
B_{10}=&0,
&B_{20}=&\frac{1}{12}(4m^2+\vq^2)B,
&B_{22}=&-\frac{1}{12}(4m^2+\vq^2)B,
\\\nonumber
B_{11}=&-\frac12 B,
&B_{21}=&0,
&B_{23}=&\frac{1}{3\vq^2}(m^2+\vq^2)B.
\end{align}
For the triangle integrals:
\begin{align}
\label{eq:miqreli}
I_{10}=&-B,
&I_{22}=&-\frac{1}{8}(4m^2+\vq^2)I,
&I_{32}=&\frac{1}{12}(4m^2+\vq^2)B,
\\\nonumber
I_{11}=&-\frac12 I,
&I_{23}=&\frac{1}{8\vq^2}(4m^2+3\vq^2)I,
&I_{33}=&-\frac{1}{3\vq^2}(m^2+\vq^2)B,
\\\nonumber
I_{20}=&0,
&I_{30}=&-\frac{1}{12}(4m^2+\vq^2)B,
&I_{34}=&\frac{1}{16}(4m^2+\vq^2)I,
\\\nonumber
I_{21}=&\frac12 B,
&I_{31}=&0.
&
\end{align}
And for the crossed-box integrals:
\begin{align}
\label{eq:miqrelj}
J_{10}=&-I,
&J_{35}=&\frac{3}{4\vq^2}B-\frac{1}{16\vq^2}(12m^2+5\vq^2)J,
\\\nonumber
J_{11}=&-\frac12 J,
&J_{40}=&\frac{4m^2+\vq^2}{12}B,
\\\nonumber
J_{20}=&B,
&J_{41}=&0,
\\\nonumber
J_{21}=&\frac12 I,
&J_{42}=&-\frac{4m^2+\vq^2}{12}B,
\\\nonumber
J_{22}=&\frac12 B-\frac{4m^2+\vq^2}{8}J,
&J_{43}=&\frac{m^2+\vq^2}{3\vq^2}B,
\\\nonumber
J_{23}=&-\frac{1}{2\vq^2}B+\frac{4m^2+3\vq^2}{8\vq^2}J,
&J_{44}=&-\frac{4m^2+\vq^2}{16}I,
\\\nonumber
J_{30}=&0,
&J_{45}=&\frac{12m^2+5\vq^2}{16\vq^2}I,
\\\nonumber
J_{31}=&-\frac12B,
&J_{46}=&-\frac{20m^2+5\vq^2}{96}B+\frac{4m^2+\vq^2}{128}J,
\\\nonumber
J_{32}=&\frac{4m^2+\vq^2}{8}I,
&J_{47}=&\frac{20m^2+17\vq^2}{96\vq^2}B
-\frac{24m^2\vq^2+16m^4+5\vq^4}{128\vq^2}J,
\\\nonumber
J_{33}=&-\frac{4m^2+3\vq^2}{8\vq^2}I,
&J_{48}=&-\frac{20m^2+29\vq^2}{32\vq^4}B
+\frac{48m^4+120m^2\vq^2+35\vq^4}{128\vq^4}J.
\\\nonumber
J_{34}=&-\frac14 B+\frac{4m^2+\vq^2}{16}J, & &
\end{align}
The relations for the integrals $K_{\mu...\nu}$ appearing in the crossed-box diagrams
can be obtained from the ones for $J_{\mu...\nu}$.
This is done by changing the sign of all the terms that do not contain $J$
and by making the replacement $J\to K$ in all the others.
In all these expressions we have neglected the constant terms
since they will not contribute to the Fourier transforms.

Using these relations and neglecting $q_0$ and $q_0'$ in
the potentials in momentum space of Ref.~\cite{axel2} we obtain
the following expressions for each of the nine loop diagrams
of Fig.~\ref{fig:nlodiagrams},
\begin{align}
\label{eq:potdiagqnoq0}
V_a=&
\frac13 g_a(4m^2+\vq^2)B\,\vtu\cdot\vtd,
\\\nonumber
V_b=&
-\frac12 g_b(2m^2+\vq^2)I,
\\\nonumber
V_c=&\frac16 g_c(8m^2+5\vq^2)B\,\vtu\cdot\vtd,
\\\nonumber
V_d=&
g_d\left[
2A_\Lambda M_N B\, \vsu\cdot\vq
-(2m^2+\vq^2)B_\Lambda B
+B_\Lambda B \vq\cdot\vp
+B_\Lambda B i\vsu\cdot(\vq\times\vp)
\right]\vtu\cdot\vtd,
\\\nonumber
V_e=&
g_e\left[B_{\Sigma_1}(2m^2+\vq^2)
-2A_{\Sigma_1}M_N\, \vsu\cdot\vq
\right]B,
\\\nonumber
V_f=&
\frac18 g_f\Big\{
A_\Lambda\left[
M_N(8B+4(2m^2+\vq^2)K)\vsu\cdot\vq
+M_N(8B+2(4m^2+\vq^2)K\,i(\vsu\times\vsd)\cdot\vq\right]
\\&\nonumber
-B_\Lambda\left[
4(2m^2+\vq^2)B+2(2m^2+\vq^2)^2K\right]
\\&\nonumber
+B_\Lambda(4B+2(2m^2+\vq^2)K)\left[
\,i\vsu\cdot(\vq\times\vp)+\vq\cdot\vp\right]
\\&\nonumber
+B_\Lambda(4B+(4m^2+\vq^2)K)\big[
i\vsd\cdot(\vq\times\vp)
+\,(\vsu\cdot\vq)(\vsd\cdot\vq)
-\vq^2\,\,(\vsu\cdot\vsd)
+\vq\cdot\vp
\\&\nonumber
-\,(\vsu\cdot\vq)(\vsd\cdot\vp)\big]
\Big\}(3-2\vtu\cdot\vtd),
\\\nonumber
V_g=&
\frac18 g_g\Big\{
A_{\Sigma_2}M_N(8B+4(2m^2+\vq^2)K)\,\vsu\cdot\vq
-A_{\Sigma_2}M_N(8B+2(4m^2+\vq^2)K)\,i(\vsu\times\vsd)\cdot\vq
\\&\nonumber
-B_{\Sigma_2}\Big[4(2m^2+\vq^2)B+2(2m^2+\vq^2)^2K\Big]
\\\nonumber &
+B_{\Sigma_2}(4B+(4m^2+\vq^2)K)\Big[
(\vsu\cdot\vq)(\vsd\cdot\vq)-\vq^2(\vsu\cdot\vsd)\Big]
\Big\},
\\\nonumber
V_h=&
\frac18 g_h\Big\{
A_\Lambda M_N(8B-4(2m^2+\vq^2)J)\vsu\cdot\vq
+A_\Lambda M_N(-8B+2(4m^2+\vq^2)J)\,i(\vsu\times\vsd)\cdot\vq
\\&\nonumber
+4B_\Lambda (2m^2-\vq^2)B+2B_\Lambda (-4m^2+6m^2\vq^2+q^4)J
\\&\nonumber
+B_\Lambda (4B-2(2m^2+\vq^2)J)\Big[\vq\cdot\vp+i\vsu\cdot(\vq\times\vp)\Big]
\\&\nonumber
+B_\Lambda (4B-(4m^2+\vq^2)J)
\\&\nonumber\times
\Big[
(\vsu\cdot\vq)(\vsd\cdot\vp)-i\,\vsd\cdot(\vq\times\vp)
-(\vsu\cdot\vq)(\vsd\cdot\vq)+\vq^2\vsu\cdot\vsd-\vq\cdot\vp\Big]
\Big\}
(3+2\vtu\cdot\vtd),
\\\nonumber
V_i=&
\frac18 g_i\Big\{
A_{\Sigma_3}M_N(-8B+4(2m^2+\vq^2)J)\vsu\cdot\vq
\\&\nonumber
+A_{\Sigma_3}M_N(-8B+2(4m^2+\vq^2)J)\,i(\vsu\times\vsd)\cdot\vq
+B_{\Sigma_3}\Big[4(2m^2+\vq^2)B-2(2m^2+\vq^2)^2J\Big]
\\&\nonumber
+B_{\Sigma_3}(4B-(4m^2+\vq^2)J)
\Big[(\vsu\cdot\vq)(\vsu\cdot\vq)-\vq^2\vsu\cdot\vsd\Big]
\Big\}.
\end{align}
We have used the same labels as in Fig.~\ref{fig:nlodiagrams}.
The strong and weak couplings are encapsulated
in the global factors $g_\alpha$ ($\alpha=a,\dots,i$), the constants
$A_\Lambda$, $B_\Lambda$, and the isospin operators
$A_{\Sigma_i}\equiv A_{\Sigma_i}^v+A_{\Sigma_i}^w\vtu\cdot\vtd$
and $B_{\Sigma_i}\equiv B_{\Sigma_i}^v+B_{\Sigma_i}^w\vtu\cdot\vtd$
($i=1,2,3$). 
They are defined as
\begin{align}
\label{eq:couplingsg}
g_a=&\frac{h_{\Lambda N}}{8f_\pi^4(M_\Lambda-M_N)},
&g_d=&-\frac{g_{NN\pi}}{8M_N^2 f_\pi^2},
&g_g=&-\frac{g_{NN\pi}^2g_{\Lambda\Sigma\pi}}{8M_N^3(M_\Sigma+M_\Lambda)},
\\\nonumber
g_b=&\frac{3h_{2\pi}g_{NN\pi}^2}{4M_N^2 f_\pi^2},
&g_e=&-\frac{g_{\Lambda\Sigma\pi}}{8M_N(M_\Sigma+M_\Lambda)f_\pi^2},
&g_h=&g_f=-\frac{g_{NN\pi}^3}{16M_N^4},
\\\nonumber
g_c=&-\frac{h_{\Lambda N}g_{NN\pi}^2}{8M_N^2f_\pi^2(M_\Lambda-M_N)},
&g_f=&-\frac{g_{NN\pi}^3}{16M_N^4},
&g_i=&-g_g=\frac{g_{NN\pi}^2g_{\Lambda\Sigma\pi}}{8M_N^3(M_\Sigma+M_\Lambda)},
\end{align}
and
\begin{align}
\label{eq:couplingsab}
A_{\Sigma_1}^v=&0,
&A_{\Sigma_2}^v=&-\sqrt{3}A_{\Sigma\frac12}+2A_{\Sigma\frac32},
&A_{\Sigma_3}^v=&-\sqrt{3}A_{\Sigma\frac12}+2A_{\Sigma\frac32},
\\\nonumber
A_{\Sigma_1}^w=&\frac23(\sqrt{3}A_{\Sigma\frac12}+A_{\Sigma\frac32}),
&A_{\Sigma_2}^w=&\frac23(\sqrt{3}A_{\Sigma\frac12}+A_{\Sigma\frac32}),
&A_{\Sigma_3}^w=&-\frac23(\sqrt{3}A_{\Sigma\frac12}+2A_{\Sigma\frac32}),
\\\nonumber
B_{\Sigma_1}^v=&0,
&B_{\Sigma_2}^v=&-\sqrt{3}B_{\Sigma\frac12}+2B_{\Sigma\frac32},
&B_{\Sigma_3}^v=&-\sqrt{3}B_{\Sigma\frac12}+2B_{\Sigma\frac32},
\\\nonumber
B_{\Sigma_1}^w=&\frac23(\sqrt{3}B_{\Sigma\frac12}+B_{\Sigma\frac32}),
&B_{\Sigma_2}^w=&\frac23(\sqrt{3}B_{\Sigma\frac12}+B_{\Sigma\frac32}),
&B_{\Sigma_3}^w=&-\frac23(\sqrt{3}B_{\Sigma\frac12}+2B_{\Sigma\frac32}).
\end{align}
The mass of the $\Sigma$ is $M_\Sigma=1193$ MeV and
the pion decay constant $f_\pi=92.4$ MeV.
$g_{\Lambda\Sigma\pi}=9.32$ is the strong coupling
for the $\Lambda \Sigma\pi$ vertex and
is taken from the Nijmegen 97f model \cite{nij99}.
The weak couplings
$A_{\Sigma\frac12}=-0.59$, $B_{\Sigma\frac12}=-15.68$,
$A_{\Sigma\frac32}=2.00$ and $B_{\Sigma\frac32}=-0.26$
are fixed by the weak decay of the $\Sigma$,
while
$h_{\Lambda N}=-(D+3F)/(\sqrt{6}G_Fm^2)=81.02$ MeV
and $h_{2\pi}=(D+3F)/(8\sqrt{6}G_Fm^2)=-10.13$ MeV
depend on $D$ and $F$, which parametrize the
weak chiral SU(3) Lagrangian.
These parameters can be fitted using the pole model
to hyperon decay data \cite{donoghue}.

\subsection{Potential in coordinate space}
\label{app:noq0potr}
The potentials derived in the previous section depend on 
the four master integrals $B$, $I$, $J$, and $K$
---which in turn can be expressed in terms of
$A(q)$, $L(q)$ or $\frac{1}{4m^2+\vq^2}L(q)$---,
and powers of momentum $\vq^{2n}$, where $n=0,1,2$.
They also include various operators with
zero, one, or two momenta $\vq$, as for example
$\vsu\cdot\vsd$, $\vsu\cdot\vq$ and
$(\vsu\cdot\vq)(\vsd\cdot\vq)$.
We call these type of operators scalar, vector, and tensor.
The Fourier transforms of the scalar functions,
\begin{align}
\label{eq:scalarfuncs}
A(q)\vq^{2n},~L(q)\vq^{2n},~\frac{L(q)}{4m^2+q^2}\vq^{2n},
\end{align}
are calculated using dispersion relations \cite{kaiser}
and their results are shown below.
\begin{align}
\label{eq:ftaqlq}
\mathcal{F}\left[A(q)\right]&=
\frac{e^{-2 m r}}{8 \pi r^2},
\\\nonumber
\mathcal{F}\left[A(q)\vec{q}\,^2\right]&=
-\frac{e^{-2 m r}}{4 \pi r^4}(1+2m r(1+m r)),
\\\nonumber
\mathcal{F}\left[A(q)\vec{q}\,^4\right]&=
\frac{e^{-2 m r}}{\pi r^6}
(3+6mr+6m^2r^2+4m^3r^3+2m^4r^4),
\\\nonumber
\mathcal{F}\left[L(q)\right]&=
-\frac{m}{2\pi r^2} K_1(2mr),
\\\nonumber
\mathcal{F}\left[L(q)\vec{q}\,^2\right]&=
\frac{m}{\pi r^4}
(3m r K_0(2mr)+(3+2m^2r^2)K_1(2mr)),
\\\nonumber
\mathcal{F}\left[L(q)\vec{q}\,^4\,\right]&=
-\frac{2m}{\pi r^6}
[6mr(5+2m^2r^2) K_0(2mr)+(30+27m^2r^2+4m^4r^4)K_1(2mr)],
\\\nonumber
\mathcal{F}\left[\frac{L(q)}{4m^2+q^2}\right]&=
\frac{1}{4\pi r} K_0(2mr),
\\\nonumber
\mathcal{F}\left[\frac{L(q)}{4m^2+q^2}\vec{q}\,^2\right]&=
-\frac{m}{2\pi r^2}
[2mr K_0(2mr)+K_1(2mr)],
\\\nonumber
\mathcal{F}\left[\frac{L(q)}{4m^2+q^2}\vec{q}\,^4\,\right]&=
\frac{m(3+4m^2r^2)}{\pi r^4}
[mr K_0(2mr)+K_1(2mr)].
\end{align}
$K_0$ and $K_1$ are the modified Bessel functions of the second kind
of zeroth and first order.
The Fourier transforms of the vector and tensor terms
can be calculated analogously to the ones for the OPE and OKE potentials.
This is done by making the replacement $\vq\to -i\vec{\nabla}$
and applying the gradients to the previous results.
The term $(\vsu\cdot\hr)(\vsd\cdot\hr)$ obtained from
the Fourier transform of $(\vsu\cdot\vq)(\vsd\cdot\vq)$
is expressed as
$(\vsu\cdot\hr)(\vsd\cdot\hr)\equiv \frac13\hat{S}_{12}(\hat{r})-\frac13\vsu\cdot\vsd$.
Therefore the spin-spin potential in coordinate space
also gets a contribution from the $(\vsu\cdot\vq)(\vsd\cdot\vq)$ terms.

We apply these relations to the potentials in momentum space
of Eq.~(\ref{eq:potdiagqnoq0}) and reorganize them in terms of the operators
of Eq.~(\ref{eq:nloops}).
Diagrams $f$ and $g$ depend on $q_0'$ in order to avoid the
pinch singularity coming from the baryonic propagators.
For diagram $f$ we have $q_0'=M_N-M_\Lambda$ and for diagram $g$
$q_0'=M_\Sigma-M_\Lambda$.
Defining $\Delta_1\equiv \frac12 (M_\Lambda-M_N)$,
$\Delta_2\equiv M_\Sigma-M_\Lambda$, and 
$\mu\equiv m r$, we obtain the following expressions
for the central, vector, and tensor potentials.

\subsubsection*{Central terms}

\begin{align}
V_1=&
\frac{1}{32 \pi ^3 r^6}\Big[\pi  \left(\left(3 B_{\Lambda } g_f \Delta_1^{-1}-2 B_{\text{$\Sigma $2}}^v g_g \Delta_2^{-1}\right)
\left(6+12 \mu +10 \mu ^2+4 \mu ^3+\mu ^4\right)
\right.\\&\nonumber\left.
-g_b (1+\mu )^2 r^2\right)e^{-2 \mu } 
 \\\nonumber & 
-2 \left(6 B_{\text{$\Sigma $1}}^v g_e+6 B_{\Lambda } g_f \left(-9+2 \mu ^2\right)-g_g \left(B_{\text{$\Sigma $2}}^v+B_{\text{$\Sigma $3}}^v\right) \left(9+2 \mu ^2\right)\right) r \mu ^2\,K_0\text{(2$\mu $)}
 \\\nonumber & 
-2 \left(2 B_{\text{$\Sigma $1}}^v g_e \left(3+\mu ^2\right)-\left(6 B_{\Lambda } g_f+g_g \left(B_{\text{$\Sigma $2}}^v+B_{\text{$\Sigma $3}}^v\right)\right) \left(9+5 \mu ^2\right)\right) r \mu\, K_1\text{(2$\mu $)}\Big],
\end{align}

\begin{align}
W_1=&
\frac{1}{48 \pi ^3 r^6}\Big[
-3 \left(2 g_a+5 g_c-6 B_{\Lambda } g_d+6 B_{\text{$\Sigma $1}}^w g_e-9 B_{\text{$\Sigma $1}}^w g_g-9 B_{\text{$\Sigma $2}}^w g_g
\right. \\\nonumber & \left.
+16 B_{\Lambda } g_f \mu ^2-2 B_{\text{$\Sigma $1}}^w g_g \mu ^2-2 B_{\text{$\Sigma $2}}^w g_g \mu^2 \right) r \mu ^2
\,K_0\text{(2$\mu $)} 
 \\\nonumber & 
-3 \left(2 g_a-6 B_{\Lambda } g_d+6 B_{\text{$\Sigma $1}}^w g_e-9 B_{\text{$\Sigma $1}}^w g_g-9 B_{\text{$\Sigma $2}}^w g_g+g_c \left(5+2 \mu ^2\right)-2 B_{\Lambda } g_d \mu ^2
\right. \\\nonumber & \left.
+2 B_{\text{$\Sigma $1}}^w g_e \mu ^2-5 B_{\text{$\Sigma $1}}^w g_g \mu ^2-5 B_{\text{$\Sigma $2}}^w g_g \mu ^2\right) r \mu \,
K_1\text{(2$\mu $)}
 \\\nonumber & 
-3 \pi  \left(B_{\Lambda } g_f \Delta_1^{-1}+B_{\text{$\Sigma $2}}^w g_g \Delta_2^{-1}\right) \left(6+12 \mu +10 \mu ^2+4 \mu ^3+\mu ^4\right)\,e^{-2 \mu }\Big],
\end{align}

\begin{align}
V_{\vsu\cdot\vsd}=&
\frac{1}{48 \pi ^3 r^6}\Big[
\pi  \left(3 B_{\Lambda } g_f \Delta_1^{-1}-2 B_{\text{$\Sigma $2}}^v g_g \Delta_2^{-1}\right) \left(3+6 \mu +5 \mu ^2+2 \mu ^3\right)
\,e^{-2 \mu } 
 \\\nonumber & 
+ 12 g_g \left(B_{\text{$\Sigma $2}}^v-B_{\text{$\Sigma $3}}^v\right) r \mu ^2\,K_0\text{(2$\mu $)}
+ 4 g_g \left(B_{\text{$\Sigma $2}}^v-B_{\text{$\Sigma $3}}^v\right) \left(3+2 \mu ^2\right) r \mu K_1\text{(2$\mu $)}\Big],
\end{align}

\begin{align}
W_{\vsu\cdot\vsd}=&
\frac{1}{24 \pi ^3 r^6}\Big[
 -\pi  \left(B_{\Lambda } g_f \Delta_1^{-1}+B_{\text{$\Sigma $2}}^w g_g \Delta_2^{-1}\right) \left(3+6 \mu +5 \mu ^2+2 \mu ^3\right)
 \,e^{-2 \mu }
 \\\nonumber & 
-6 \left(4 B_{\Lambda } g_f+g_g \left(B_{\text{$\Sigma $1}}^w-B_{\text{$\Sigma $2}}^w\right)\right) r \mu ^2
\,K_0\text{(2$\mu $)} 
\\\nonumber&
-2 \left(4 B_{\Lambda } g_f+g_g \left(B_{\text{$\Sigma $1}}^w
-B_{\text{$\Sigma $2}}^w\right)\right) \left(3+2 \mu ^2\right) r \mu
\,K_1\text{(2$\mu $)} \Big].
\end{align}


\subsubsection*{Vector terms}

\begin{align}
V_{\vsu\cdot\hr}=&
\frac{i}{16 \pi ^3 r^5}\Big[
 -4  M_N \left(A_{\text{$\Sigma $1}}^v g_e-2 \left(6 A_{\Lambda } g_f+g_g \left(A_{\text{$\Sigma $2}}^v+A_{\text{$\Sigma $3}}^v\right)\right)\right) r \mu ^2\,K_0\text{(2$\mu $)}
 \\\nonumber & 
+  M_N \left(-6 A_{\text{$\Sigma $1}}^v g_e+\left(6 A_{\Lambda } g_f+g_g \left(A_{\text{$\Sigma $2}}^v+A_{\text{$\Sigma $3}}^v\right)\right) \left(9+4 \mu ^2\right)\right) r \mu \, K_1\text{(2$\mu $)}
 \\\nonumber & 
+\pi  M_N \left(3 A_{\Lambda } g_f \Delta_1^{-1}-2 A_{\text{$\Sigma $2}}^v g_g \Delta_2^{-1}\right) \left(2+4 \mu +3 \mu ^2+\mu ^3\right)\,e^{-2 \mu } 
\Big],
\end{align}

\begin{align}
W_{\vsu\cdot\hr}=&
\frac{i}{16 \pi ^3 r^5}\Big[
 4  M_N \left(A_{\Lambda } g_d-A_{\text{$\Sigma $1}}^w g_e+2 g_g \left(A_{\text{$\Sigma $2}}^w+A_{\text{$\Sigma $3}}^w\right)\right) r \mu ^2
 \,K_0\text{(2$\mu $)}
 \\\nonumber & 
+ M_N \left(6 A_{\Lambda } g_d-6 A_{\text{$\Sigma $1}}^w g_e+g_g \left(A_{\text{$\Sigma $2}}^w+A_{\text{$\Sigma $3}}^w\right) \left(9+4 \mu ^2\right)\right) r \mu
\,K_1\text{(2$\mu $)} 
\\\nonumber &
-2 i \pi  M_N \left(A_{\Lambda } g_f \Delta_1^{-1}+A_{\text{$\Sigma $2}}^w g_g \Delta_2^{-1}\right) \left(2+4 \mu +3 \mu ^2+\mu ^3\right)
\,e^{-2 \mu }
\Big],
\end{align}

\begin{align}
V_{\hr\cdot\hp}=&
\frac{i}{32 \pi ^3 r^6}\Big[
3 \pi  B_{\Lambda } g_f \Delta_1^{-1} \left(2+4 \mu +3 \mu ^2+\mu ^3\right)\,e^{-2 \mu }
+ 48  B_{\Lambda } g_f r \mu ^2 K_0\text{(2$\mu $)}
\\\nonumber &
+ 6  B_{\Lambda } g_f \left(9+4 \mu ^2\right) r \mu \,K_1\text{(2$\mu $)}\Big],
\end{align}

\begin{align}
W_{\hr\cdot\hp}=&
\frac{i}{16 \pi ^3 r^6}\Big[
2  B_{\Lambda } g_d r \mu ^2\,K_0\text{(2$\mu $)} 
+ 3  B_{\Lambda } g_d r \mu \,K_1\text{(2$\mu $)}
\\\nonumber&
-\pi  B_{\Lambda } g_f \Delta_1^{-1} \left(2+4 \mu +3 \mu ^2+\mu ^3\right)\,e^{-2 \mu } 
\Big],
\end{align}

\begin{align}
V_{(\vsu\cdot\vsd)(\hr\cdot\hp)}=&
\frac{3i}{32 \pi ^2 r^6}
 B_{\Lambda } g_f \Delta_1^{-1} (1+\mu )^2\,e^{-2 \mu },
\end{align}

\begin{align}
W_{(\vsu\cdot\vsd)(\hr\cdot\hp)}=&
\frac{-i}{16 \pi ^3 r^6}\Big[
\pi  B_{\Lambda } g_f \Delta_1^{-1} (1+\mu )^2
\,e^{-2 \mu }
 +8  B_{\Lambda } g_f r \mu ^2\,K_0\text{(2$\mu $)}
 +12  B_{\Lambda } g_f r \mu \,K_1\text{(2$\mu $)} 
 \Big],
\end{align}

\begin{align}
V_{(\vsu\times\vsd)\cdot\hr}=&
\frac{1}{16 \pi ^3 r^5}\Big[
-\pi  M_N \left(3 A_{\Lambda } g_f \Delta_1^{-1}+2 A_{\text{$\Sigma $2}}^v g_g \Delta_2^{-1}\right) (1+\mu )^2\,e^{-2 \mu }
 \\\nonumber & 
+ 4 g_g M_N \left(A_{\text{$\Sigma $2}}^v-A_{\text{$\Sigma $3}}^v\right) r \mu ^2\,K_0\text{(2$\mu $)}
+6 g_g M_N \left(A_{\text{$\Sigma $2}}^v-A_{\text{$\Sigma $3}}^v\right) r \mu \,K_1\text{(2$\mu $)} 
\Big],
\end{align}

\begin{align}
W_{(\vsu\times\vsd)\cdot\hr}=&
\frac{1}{8 \pi ^3 r^5}\Big[
\pi  M_N \left(A_{\Lambda } g_f \Delta_1^{-1}-A_{\text{$\Sigma $2}}^w g_g \Delta_2^{-1}\right) (1+\mu )^2\,e^{-2 \mu }
 \\\nonumber & 
+ 2 M_N \left(4 A_{\Lambda } g_f+g_g \left(A_{\text{$\Sigma $2}}^w-A_{\text{$\Sigma $3}}^w\right)\right) r \mu ^2\,K_0\text{(2$\mu $)}
 \\\nonumber & 
+ 3 M_N \left(4 A_{\Lambda } g_f+g_g \left(A_{\text{$\Sigma $2}}^w-A_{\text{$\Sigma $3}}^w\right)\right) r \mu \,K_1\text{(2$\mu $)}
\Big],
\end{align}

\begin{align}
V_{\vsu\cdot(\hr\times\hp)}=&
\frac{1}{32 \pi ^3 r^6}\Big[
-3 \pi  B_{\Lambda } g_f \Delta_1^{-1} \left(2+4 \mu +3 \mu ^2+\mu ^3\right)\,e^{-2 \mu }
-48 B_{\Lambda } g_f r \mu ^2\,K_0\text{(2$\mu $)}
\\\nonumber&
-6 B_{\Lambda } g_f \left(9+4 \mu ^2\right) r \mu \,K_1\text{(2$\mu $)}\Big],
\end{align}

\begin{align}
W_{\vsu\cdot(\hr\times\hp)}=&
\frac{1}{16 \pi ^3 r^6}\Big[
 -2 B_{\Lambda } g_d r \mu ^2\,K_0\text{(2$\mu $)}
-3 B_{\Lambda } g_d r \mu \,K_1\text{(2$\mu $)}
\\\nonumber &
+ \pi  B_{\Lambda } g_f \Delta_1^{-1} \left(2+4 \mu +3 \mu ^2+\mu ^3\right)\,e^{-2 \mu }
\Big],
\end{align}

\begin{align}
V_{\vsd\cdot(\hr\times\hp)}=&
\frac{-3}{32 \pi ^2 r^6}
 B_{\Lambda } g_f \Delta_1^{-1} (1+\mu )^2\,e^{-2 \mu },
\end{align}

\begin{align}
W_{\vsd\cdot(\hr\times\hp)}=&
\frac{1}{16 \pi ^3 r^6}\Big[
\pi  B_{\Lambda } g_f \Delta_1^{-1} (1+\mu )^2\,e^{-2 \mu }
+ 8 B_{\Lambda } g_f r \mu ^2\,K_0\text{(2$\mu $)}
+12 B_{\Lambda } g_f r \mu \,K_1\text{(2$\mu $)} 
\Big],
\end{align}

\begin{align}
V_{(\vsu\cdot\hr)(\vsd\cdot\hp)}=&
\frac{-3i}{32 \pi ^2 r^6}
  B_{\Lambda } g_f \Delta_1^{-1} (1+\mu )^2\,e^{-2 \mu },
\end{align}

\begin{align}
W_{(\vsu\cdot\hr)(\vsd\cdot\hp)}=&
\frac{i}{16 \pi ^3 r^6}\Big[
 \pi  B_{\Lambda } g_f \Delta_1^{-1} (1+\mu )^2\,e^{-2 \mu } 
 + 8  B_{\Lambda } g_f r \mu ^2\,K_0\text{(2$\mu $)}
 + 12  B_{\Lambda } g_f r \mu \,K_1\text{(2$\mu $)}
 \Big].
\end{align}

\subsubsection*{Tensor terms}

\begin{align}
V_{\hat{S}_{12}}=&
\frac{1}{48 \pi ^3 r^6}\Big[
-\pi\left(3 B_{\Lambda } g_f \Delta_1^{-1}-2 B_{\text{$\Sigma $2}}^v g_g \Delta_2^{-1}\right) \left(3+6 \mu +4 \mu ^2+\mu ^3\right)\,e^{-2 \mu }
 \\\nonumber & 
-12 g_g \left(B_{\text{$\Sigma $2}}^v-B_{\text{$\Sigma $3}}^v r\right) \mu ^2\,K_0\text{(2$\mu $)}
-g_g \left(B_{\text{$\Sigma $2}}^v-B_{\text{$\Sigma $3}}^v\right) \left(15+4 \mu ^2\right) r \mu \,K_1\text{(2$\mu $)}
\Big],
\end{align}

\begin{align}
W_{\hat{S}_{12}}=&
\frac{1}{48 \pi ^3 r^6}\Big[
 2 \pi  \left(B_{\Lambda } g_f \Delta_1^{-1}+B_{\text{$\Sigma $2}}^w g_g \Delta_2^{-1}\right) \left(3+6 \mu +4 \mu ^2+\mu ^3\right)
 \,e^{-2 \mu }
 \\\nonumber & 
+ 12 \left(4 B_{\Lambda } g_f+g_g \left(B_{\text{$\Sigma $1}}^w-B_{\text{$\Sigma $2}}^w\right)\right) r \mu ^2 K_0\text{(2$\mu $)}
\\\nonumber&
+ \left(4 B_{\Lambda } g_f+g_g \left(B_{\text{$\Sigma $1}}^w-B_{\text{$\Sigma $2}}^w\right)\right) \left(15+4 \mu ^2\right) r \mu
\,K_1\text{(2$\mu $)}\Big].
\end{align}


\section{Fourier transform of the potential for $q_0\neq0$ and $q_0'\neq0$}
\label{app:q0}

When the baryonic mass differences are explicitly considered 
it is simpler to calculate the potentials directly in coordinate space.
Instead of calculating the master integrals in momentum space
and then Fourier transforming them,
we rewrite the loop diagrams in terms of the coordinate integrals
of Eq.~(\ref{eq:mirq0}).
In the first section of this Appendix we derive these integrals
and show their results in terms of a simple set of functions.
The $\Lambda N\to NN$ potentials for each diagram and for each operator
are shown in terms of these functions in the second section.

\subsection{Master integrals in coordinate space}
\label{app:q0mir}
As in Eq.~(\ref{eq:miqorg}), the integrals appearing in the potentials
in coordinate space are organized into
structures of Kronecker deltas and vectors $\vr$.
In this Appendix we denote the integrals in coordinate space
with the same letters as the ones used for the momentum ones
in the previous Appendix.
We show only the case for the integrals $J$,
but the same definitions can be applied to the others.
\begin{align}
\label{eq:mirorg}
J_\mu\equiv\,&
\delta_{\mu0}J_{10}+\delta_{\mu
i}J_{11}\vr_i \,,
\\\nonumber
J_{\mu\nu}\equiv\,&
\delta_{\mu0}\delta_{\nu0}J_{20}
+(\delta_{\mu0}\delta_{\nu i}
+\delta_{\mu i}\delta_{\nu 0})J_{21}\vr_i
+\delta_{\mu i}\delta_{\nu j}(J_{22}\delta_{ij}
+J_{23}\vr_i\vr_j)\,,
\\\nonumber
J_{\mu\nu\rho}\equiv\,&
\delta_{\mu0}\delta_{\nu0}\delta_{\rho0}J_{30}
+\delta\delta\delta_{\{\mu\nu\rho 00i\}}\vr_iJ_{31}
+\delta\delta\delta_{\{\mu\nu\rho 0ij\}}
(\delta_{ij}J_{32}+\vr_i\vr_jJ_{33})
\\&
+\delta_{\mu i}\delta_{\nu j}\delta_{\rho k}
(\delta\vr_{\{ijk\}}J_{34}+\vr_i\vr_j\vr_kJ_{35})\,,
\nonumber\\\nonumber
J_{\mu\nu\rho\sigma}\equiv\,&
\delta_{\mu0}\delta_{\nu0}\delta_{\rho0}\delta_{\sigma0}J_{40}
+\delta\delta\delta\delta_{\{\mu\nu\rho\sigma000i\}}\vr_iJ_{41}
+\delta\delta\delta\delta_{\{\mu\nu\rho\sigma00ij\}}
(\delta_{ij}J_{42}+\vr_i\vr_jJ_{43})
\\&
+\delta\delta\delta\delta_{\{\mu\nu\rho\sigma0ijk\}}
(\delta\vr_{\{ijk\}}J_{44}+\vr_i\vr_j\vr_kJ_{45})
\nonumber\\&
+\delta_{\mu i}\delta_{\nu j}\delta_{\rho k}\delta_{\sigma l}
(\delta\delta_{\{ijkl\}}J_{46}
+\delta\vr\vr_{\{ijkl\}}J_{47}
+\vr_i\vr_j\vr_k\vr_lJ_{48})\,. \nonumber
\end{align}
We have used the following definitions:
\begin{align}
\label{eq:deltarstr}
\delta\vr_{\{ijk\}}=&\,
\delta_{ij}\vr_k
+\delta_{ik}\vr_j
+\delta_{jk}\vr_i\,,
\\\nonumber
\delta\vr\vr_{\{ijkl\}}=&\,
\delta_{ij}\vr_k\vr_l
+\delta_{ik}\vr_j\vr_l
+\delta_{il}\vr_j\vr_k 
+\delta_{jk}\vr_i\vr_l
+\delta_{jl}\vr_i\vr_k
+\delta_{kl}\vr_i\vr_j\,,
\\\nonumber
\delta\delta_{\{ijkl\}}=&\,
\delta_{ij}\delta_{kl}
+\delta_{ik}\delta_{jl}
+\delta_{il}\delta_{jk}\,.
\end{align}
First we calculate the integrals $B$ and second 
the integrals $I$, $J$, and $K$.

\subsubsection*{$B$ integrals}
We use the Veltman-Passarino reductions for the ball integrals
in momentum space and Fourier-transform them.
This allows us to write $B$, $B_\mu$, and $B_{\mu\nu}$
in terms of only two integrals, $b_1$ and $b_2$,
and their gradients:
\begin{align}
\label{eq:relbr}
B=&b_1
\\\nonumber
B_0=&-\frac{q_0}{2}b_1,
\\\nonumber
B_i=&
\frac{i}{2}\vec{\nabla}_i b_1,
\\\nonumber
B_{00}=&\frac{1}{12}(4m^2+3q_0^2-\vec{\nabla}^2)b_1-\frac{m^2}{3}b_2,
\\\nonumber
B_{0i}=&-i\frac{q_0}{3}\vec{\nabla}_i (b_1-m^2b_2),
\\\nonumber
B_{ij}=&-\delta_{ij}\frac{1}{12}(4m^2-q_0^2-\vec{\nabla}^2)b_1
-\frac13\vec{\nabla}_i\vec{\nabla}_j(b_1-m^2b_2).
\end{align}
The temporal subindices are denoted with $0$ and the spatial ones
with $i$ and $j$.
Once the gradients are applied in the previous expressions
the $B$ integrals depend on the five following functions:
\begin{align}
\label{eq:bi}
b_1\equiv&
\frac{1}{32\pi^3r}
\int_{2m}^{\infty} d\nu
e^{-r\sqrt{\nu^2-q_0^2}}\sqrt{\nu^2-4m^2},
\\\nonumber
b_2\equiv&
\frac{1}{32\pi^3r}
\int_{2m}^{\infty} d\nu
e^{-r\sqrt{\nu^2-q_0^2}}\frac{\sqrt{\nu^2-4m^2}}{\nu^2},
\\\nonumber
b_3\equiv&
\frac{1}{32\pi^3r}
\int_{2m}^{\infty} d\nu
e^{-r\sqrt{\nu^2-q_0^2}} \sqrt{\nu^2-4m^2} \nu^2,
\\\nonumber
b_4\equiv&
\frac{1}{32\pi^3r}
\int_{2m}^{\infty} d\nu
e^{-r\sqrt{\nu^2-q_0^2}}\sqrt{\nu^2-4m^2}\sqrt{\nu^2-q_0^2},
\\\nonumber
b_5\equiv&
\frac{1}{32\pi^3r}
\int_{2m}^{\infty} d\nu
e^{-r\sqrt{\nu^2-q_0^2}}\frac{\sqrt{\nu^2-4m^2}}{\nu^2}\sqrt{\nu^2-q_0^2}.
\end{align}
$b_3$, $b_4$, and $b_5$ appear in the first and second derivatives
of $b_1$ and $b_2$. Specifically, these derivatives are
\begin{align}
\label{eq:bider}
b_1'=&b_4,
&b_2'=&b_5,
\\\nonumber
b_1''=&b_3-q_0^2b_1,
&b_2''=&b_1-q_0^2b_2.
\end{align}
The master integrals for each operator in terms of these
five numerical integrals are
\begin{align}
\label{eq:mibr}
B=&b_1,
\\\nonumber
B_{10}=&
-\frac{q_0}{2}b_1,
\\\nonumber
B_{11}=&
-\frac{i}{2}\left(\frac{b_1}{r^2}+\frac{b_4}{r}\right),
\\\nonumber
B_{20}=&
\frac{1}{12}
\left[4(m^2+q_0^2)b_1-4m^2q_0^2\,b_2-b_3\right],
\\\nonumber
B_{21}=&
\frac{-iq_0}{3}
\left(
-\frac{1}{r^2}b_1+\frac{m^2}{r^2}b_2-\frac{1}{r}b_4
+\frac{m^2}{r}b_5
\right),
\\\nonumber
B_{22}=&
\frac{1}{12}
\left[
4\left(\frac{1}{r^2}-m^2\right)b_1-\frac{4m^2}{r^2}b_2+b_3+\frac{4}{r}b_4
\right],
\\\nonumber
B_{23}=&
\frac{1}{12}
\left[
\left(-\frac{12}{r^4}+\frac{4q_0^2}{r^2}+\frac{4m^2}{r^2}\right)b_1
+\left(\frac{12m^2}{r^4}-\frac{4m^2q_0^2}{r^2}\right)b_2
-\frac{4}{r^2}b_3-\frac{12}{r^3}b_4+\frac{12m^2}{r^3}b_5
\right].
\end{align}

\subsubsection*{Integrals $I$, $J$, and $K$}

In order to calculate the integrals $I_{\mu...\nu}$, $J_{\mu...\nu}$,
and $K_{\mu...\nu}$,
we first make the change of variables $\vq\to\vq-\vl$.
For example, for $I_i$ we have
\begin{align}
\label{eq:iir}
I_{i}=&
\frac1i\int\frac{d^3q}{(2\pi)^3}e^{i\vec{q}\cdot\vec{r}}\int \frac{d^4l}{(2\pi)^4}
\frac{1}{l_0^2-\vl^2-m^2+i\epsilon}
\frac{1}{(l_0+q_0)^2-(\vl+\vq)^2-m^2+i\epsilon}
\frac{l_i}{-l_0-q_0'+i\epsilon}
\nonumber\\=&
\frac1i
\int\frac{dl_0}{2\pi}
\frac{1}{-l_0-q_0'+i\epsilon}
\int\frac{d^3q}{(2\pi)^3}
\frac{e^{i\vec{q}\cdot\vec{r}}}{(l_0+q_0)^2-\vq^2-m^2+i\epsilon}
\int \frac{d^3l}{(2\pi)^3}
\frac{e^{i\vl\cdot\vr}l_i}{l_0^2-\vl^2-m^2+i\epsilon}.
\end{align}
This allows us to calculate the integrals in
$\vl$ and $\vq$ separately.
Moreover the momenta $l_\mu...\l_\nu$ with spatial subindices can be replaced
by gradients that act outside the $\vl$ integral.
Therefore, the integrals in $\vl$ and $\vq$ are the same for
all $I_{i...j}$, $J_{i...j}$, and $K_{i...j}$, $i$ and $j$ denoting
spatial subindices.
When the gradients are applied to the result of the integral in $\vl$,
we obtain a sum of terms that depend on $l_0$.
This final integral in $l_0$ is calculated by residues and
depends on the various structures of Eq.~(\ref{eq:mirorg}).
We denote the coefficients accompanying these structures
$I_{rs}$, $J_{rs}$, and $K_{rs}$.
The subindex $r$ corresponds to the number of relativistic momenta in the numerator
of the integral,
and $s$ indicates the structure that the coefficient is factoring.
Their results are expressed in the following identities
as a sum of one-dimensional numerical integrals that can be
easily computed.
\begin{align}
\label{eq:irs}
I_{rs}=&
\frac{1}{16\pi^2 r}
\sum_{n=0}^{r}C_{rs}^n(r)\Bigg\{
i(-1)^n
\int_0^\infty \frac{dy}{(2\pi)}
\frac{e^{-r\left(\sqrt{m^2+(y-i\frac{q_0}{2})^2}
+\sqrt{m^2+(y+i\frac{q_0}{2})^2}\right)}}
{y^2+(\frac{q_0}{2}-q_0')^2}
\\&\nonumber\times
\left[
\left((y-i\left(\frac{q_0}{2}-q_0'\right)\right)
\sqrt{m^2+\left(y+i\frac{q_0}{2}\right)^2}^{\,n}
\right.\\\nonumber&\left.
-\left(y+i\left(\frac{q_0}{2}-q_0'\right)\right)
\sqrt{m^2+\left(y-i\frac{q_0}{2}\right)^2}^{\,n}
\right]
\\&\nonumber
+(-1)^{n+1}\theta(q_0-2q_0')
\sqrt{m^2-{q_0'}^2+i\epsilon q_0'}^n
\\&\nonumber\times
e^{-r\left(
\sqrt{m^2-(q_0-q_0')^2-i\epsilon(q_0-q_0')}
+\sqrt{m^2-{q_0'}^2+i\epsilon q_0'}
\right)}\Bigg\},
\end{align}

\begin{align}
\label{eq:jrs}
J_{rs}=&
\frac{1}{16\pi r}
\sum_{n=0}^{r}C_{rs}^n(r)
\Bigg\{
(-1)^{n+1}
\int_0^\infty \frac{dy}{(2\pi)}
\frac{e^{-r\left(\sqrt{m^2+(y-i\frac{q_0}{2})^2}
+\sqrt{m^2+(y+i\frac{q_0}{2})^2}\right)}}
{(y^2+(\frac{q_0}{2})^2)(y^2+(\frac{q_0}{2}-q_0')^2)}
\\&\nonumber\times
\left[
\left((y-i\left(\frac{q_0}{2}-q_0'\right)\right)
\left(y-i\frac{q_0}{2}\right)
\sqrt{m^2+\left(y+i\frac{q_0}{2}\right)^2}^{\,n}
\right.\\\nonumber&\left.+
\left(y+i\left(\frac{q_0}{2}-q_0'\right)\right)
\left(y+i\frac{q_0}{2}\right)
\sqrt{m^2+\left(y-i\frac{q_0}{2}\right)^2}^{\,n}
\right]
\\&\nonumber
+(-1)^{n+1}\theta(q_0-2q_0')
\frac{\sqrt{m^2-{q_0'}^2+i\epsilon q_0'}^n}{q_0'}
\\&\nonumber\times
e^{-r\left(
\sqrt{m^2-(q_0-q_0')^2-i\epsilon(q_0-q_0')}
+\sqrt{m^2-{q_0'}^2+i\epsilon q_0'}
\right)}
\\&\nonumber
+\theta(q_0)
\frac{(-m)^n}{q_0'}
e^{-r\left(
m+\sqrt{m^2-{q_0}^2-i\epsilon q_0}
\right)}\Bigg\},
\end{align}


\begin{align}
\label{eq:krs}
K_{rs}=&
J_{rs}
+\frac{1}{16\pi^2r}
\sum_{n=0}^{r}C_{rs}^n(r)
\frac{(-m)^n}{q_0'}
e^{-r\left(
m+\sqrt{m^2-{q_0}^2-i\epsilon q_0}
\right)}.
\end{align}
All the dependence on the subindices $r$ and $s$ is on the
functions $C_{rs}^{\,n}(r)$, which are given in
Table~\ref{tab:coeffscrsn}.
\begin{table}[h]
\renewcommand{\arraystretch}{1.5}
\centering
\begin{tabular}{c|c|c|c|c|c}
  & $n=0$  & $n=1$  & $n=2$  & $n=3$ & $n=4$
\\\hline
$C_{11}^{\,n}(r)$ & $-\frac{i}{r^3}$ & $\frac{i}{r^2}$ & & &
\\\hline
$C_{22}^{\,n}(r)$ & $\frac{1}{r^3}$ & $-\frac{1}{r^2}$ & & &
\\\hline
$C_{23}^{\,n}(r)$ & $-\frac{3}{r^5}$ & $\frac{3}{r^4}$ & -$\frac{1}{r^3}$ & &
\\\hline
$C_{34}^{\,n}(r)$ & $-\frac{3i}{r^5}$ & $\frac{3i}{r^4}$ & $-\frac{i}{r^3}$ & &
\\\hline
$C_{35}^{\,n}(r)$ & $\frac{15i}{r^7}$ & $-\frac{15i}{r^6}$ & $\frac{6i}{r^5}$ &
$-\frac{i}{r^4}$ &
\\\hline
$C_{46}^{\,n}(r)$ & $\frac{3}{r^5}$ & $-\frac{3}{r^4}$ & $\frac{1}{r^3}$ & &
\\\hline
$C_{47}^{\,n}(r)$ & $-\frac{15}{r^7}$ & $\frac{15}{r^6}$ & $-\frac{6}{r^5}$ &
$\frac{1}{r^4}$ &
\\\hline
$C_{48}^{\,n}(r)$ & $\frac{105}{r^9}$ & $-\frac{105}{r^8}$ & $\frac{45}{r^7}$ &
$-\frac{10}{r^6}$ & $\frac{1}{r^5}$
\end{tabular}
\caption{Functions appearing in the definitions of the master integrals.}
\label{tab:coeffscrsn}
\end{table}

The integrals $I_{rs}$, $J_{rs}$ and $K_{rs}$ with temporal subindices
can be computed from the ones
with only spatial subindices.
Using Veltman-Passarino reductions we find the following relations
\begin{align}
\label{eq:reltempr}
I_{rs}=&-B_{(r-1)s}-q_0'I_{(r-1)s},
\\\nonumber
J_{rs}=&-I_{(r-1)s},
\\\nonumber
K_{rs}=&I_{(r-1)s},
\end{align}
where $B_{00}$ and $I_{00}$ correspond to $B$ and $I$ respectively.
For example $I_{20}=-B_{10}-q_0'I_{10}$, $J_{32}=-I_{22}$,
and $K_{45}=I_{35}$.

It may also be useful to note the following relations between the integrals:
\begin{align}
\label{eq:relextrar}
I_{22}=&i I_{11},
\\\nonumber
I_{34}=&i I_{23},
\\\nonumber
J_{22}=&i J_{11},
\\\nonumber
J_{34}=&i J_{23},
\\\nonumber
J_{46}=&- J_{23},
\\\nonumber
J_{47}=&i J_{35}.
\end{align}


\subsection{Potentials}
\label{app:q0pot}
In this section we show the two-pion exchange
potentials in coordinate space for the $\Lambda N\to NN$ transition.
They are expressed in terms of the master integrals
of the previous section and their derivatives.
The expressions in momentum space of Ref.~\cite{axel2}
include vector and tensor operators, i.e. operators
with one and two momenta $\vq$.
Since these momenta are replaced by gradients,
the potentials in coordinate space will contain up to two derivatives
in the master integrals.
The potentials also depend on the mass differences $q_0$ and $q_0'$.
These quantities vary in each diagram, but all of them
can be expressed in terms of two baryonic mass differences,
namely $\Delta_1\equiv\frac12(M_\Lambda-M_N)$
and $\Delta_2\equiv M_\Sigma-M_\Lambda$.
The different values for each diagram are expressed in Table~\ref{tab:q0q0p}.
\begin{table}[htb]
\centering
\begin{tabular}{cccccccccc}
Diagram & a & b & c & d & e & f & g & h & i
\\\hline
$q_0$ & $-\Delta_1$ & $\Delta_1$ & $\Delta_1$ & $-\Delta_1$ & $-\Delta_1$
 & $-\Delta_1$ & $-\Delta_1$ & $\Delta_1$ & $\Delta_1$
\\\hline
$q_0'$ &  & $0$ & $0$ & $-2\Delta_1$ & $\Delta_2$ & $-2\Delta_1$ & $\Delta_2$
 & $-\Delta_1$ & $\Delta_1+\Delta_2$
\end{tabular}
\caption{Values of $q_0$ and $q_0'$ that appear in the integrals
of each diagram of Fig.~\ref{fig:nlodiagrams}. Note that
the ball diagram (a) only depends on $q_0$. The diagrams (b) and (c)
depend on both $q_0$ and $q_0'$ but the physical case corresponds to $q_0'=0$.}
\label{tab:q0q0p}
\end{table}

Labeling the two-pion-exchanges as in Fig.~\ref{fig:nlodiagrams} 
we obtain, for each diagram, the following expressions:
\begin{align}
V_a=&
g_a \Big[4 B_{20}+\Delta _1 \left(B \Delta _1-4 B_{10}\right)\Big]\vtu\cdot\vtd,
\end{align}
\begin{align}
V_b=&
g_b\Big[-3 i I_{11}+3 I_{22}+r \left(r I_{23}-i I_{11}'\right)\Big],
\end{align}
\begin{align}
V_c=&
g_c \Big[-6 i I_{21}+6 I_{32}+2 r^2 I_{33}-3 i \Delta _1 I_{11}+3 \Delta _1 I_{22}+\Delta _1 r^2 I_{23}-i \Delta _1 r I_{11}'-2 i r I_{21}'\Big]
\vtu\cdot\vtd,
\end{align}
\begin{align}
V_d=&
\frac{g_d }{r}\Big[B_{\Lambda } \left(i \Delta _1 r I_{11}-2 i r I_{21}+\Delta _1 I'-2 I_{10}'\right) \vsu\cdot(\vr\times\vp)
 \\\nonumber & 
+B_{\Lambda } \left(\Delta _1 r I_{11}-2 r I_{21}-i \Delta _1 I'+2 i I_{10}'\right) \vr\cdot\vp
 \\\nonumber & 
+2 A_{\Lambda } M_N \left(\Delta _1 r I_{11}-2 r I_{21}-i \Delta _1 I'+2 i I_{10}'\right) (\vsu\cdot\vr)
 \\\nonumber & 
+B_{\Lambda } r \left(-2 I \Delta _1^3+7 \Delta _1^2 I_{10}-3 i \Delta _1 I_{11}-7 \Delta _1 I_{20}+6 i I_{21}+3 \Delta _1 I_{22}
\right. \\\nonumber & \left.
+\Delta _1 r^2 I_{23}+2 I_{30}-6 I_{32}-2 r^2 I_{33}-i \Delta _1 r I_{11}'+2 i r I_{21}'\right)\Big]\vtu\cdot\vtd,
\end{align}
\begin{align}
V_e=&
g_e\Big[2 A_{\text{$\Sigma $e}} M_N \left(\Delta _1 I_{11}-2 I_{21}\right) (\vsu\cdot\vr)+B_{\text{$\Sigma $e}} \left(-2 \Delta _1^2 I_{10}+\Delta _1 \Delta _2 I_{10}+3 i \Delta _1 I_{11}+5 \Delta _1 I_{20}\right. 
\nonumber \\ & 
\left. -2 \Delta _2 I_{20}-6 i I_{21}-3 \Delta _1 I_{22}-\Delta _1 r^2 I_{23}-2 I_{30}+6 I_{32}+2 r^2 I_{33}+i \Delta _1 r I_{11}'-2 i r I_{21}'\right)\Big],
\end{align}
\begin{align}
V_f=&
\frac{-3 g_f+2 g_f \vtu\cdot\vtd}{r^3}\Big[-i B_{\Lambda } r^2 \left(r K_{23}-K_{22}'\right) (\vsu\cdot\vr) (\vsd\cdot\vp)
 \\\nonumber & 
+B_{\Lambda } r^2 \left(-r K_{23}+K_{22}'\right) \vsd\cdot(\vr\times\vp)+2 A_{\Lambda } M_N r^2 \left(-r K_{23}+K_{22}'\right) (\vsu\times\vsd)\cdot\vr
 \\\nonumber & 
+B_{\Lambda } r^2 \left(-6 i r K_{23}+5 r K_{34}+r^3 K_{35}-4 K_{11}'-4 i K_{22}'-2 i r^2 K_{23}'-r K_{11}''\right) \vr\cdot\vp
 \\\nonumber & 
+2 A_{\Lambda } M_N r^2 \left(-6 i r K_{23}+5 r K_{34}+r^3 K_{35}-4 K_{11}'-4 i K_{22}'-2 i r^2 K_{23}'-r K_{11}''\right) (\vsu\cdot\vr)
 \\\nonumber & 
+B_{\Lambda } r^2 \left(6 r K_{23}+5 i r K_{34}+i r^3 K_{35}-4 i K_{11}'+4 K_{22}'+2 r^2 K_{23}'-i r K_{11}''\right) \vsu\cdot(\vr\times\vp)
\\\nonumber &
+B_{\Lambda } \left(r \left(r K_{23}'-K_{22}''\right)+K_{22}'\right) (\vsu\cdot\vr)(\vsd\cdot\vr)
\\\nonumber &
+\left(B_{\Lambda } r^2 \left(r \left(-r K_{23}'+K_{22}''\right)-2 r K_{23}+K_{22}'\right)+i B_{\Lambda } r^2 \left(r K_{23}-K_{22}'\right) \vr\cdot\vp\right) \vsu\cdot\vsd
\\\nonumber &
+B_{\Lambda } r^2 \left(-6 i \Delta _1^2 r K_{11}+3 i \Delta _1 r K_{21}+6 \Delta _1^2 r K_{22}+2 r \left(-6+\Delta _1^2 r^2\right) K_{23}+3 i r K_{31}
\right.\\\nonumber &\left.
-\Delta _1 r^3 K_{33}-30 i r K_{34}-10 i r^3 K_{35}-3 r K_{42}-r^3 K_{43}+15 r K_{46}+10 r^3 K_{47}
\right. \\\nonumber & \left.
+r^5 K_{48}-2 i \Delta _1^2 r^2 K_{11}'+i \Delta _1 r^2 K_{21}'-2 K_{22}'-8 r^2 K_{23}'+i r^2 K_{31}'-10 i r^2 K_{34}'-2 i r^4 K_{35}'
\right.\\\nonumber & \left.
-r K_{22}''-r^3 K_{23}''\right)\Big],
\end{align}
\begin{align}
V_g=&
\frac{g_g}{r^3}\Big[2 A_{\text{$\Sigma $g}} M_N r^2 \left(-r K_{23}+K_{22}'\right) (\vsu\times\vsd)\cdot\vr
\\\nonumber & 
+2 A_{\text{$\Sigma $g}} M_N r^2 \left(-4 i r K_{23}+5 r K_{34}+r^3 K_{35}-i K_{22}'-i r^2 K_{23}'\right) (\vsu\cdot\vr)
\\\nonumber & 
+B_{\text{$\Sigma $g}} r^2 \left(r \left(r K_{23}'-K_{22}''\right)+2 r K_{23}-K_{22}'\right) \vsu\cdot\vsd
\\\nonumber & 
-B_{\text{$\Sigma $g}} \left(r \left(r K_{23}'-K_{22}''\right)+K_{22}'\right) (\vsu\cdot\vr)(\vsd\cdot\vr)
\\\nonumber & 
+B_{\text{$\Sigma $g}} r^2 \left(-3 i \Delta _2 r K_{21}+12 r K_{23}-3 i r K_{31}+3 \Delta _2 r K_{32}+\Delta _2 r^3 K_{33}+30 i r K_{34}
\right. \\\nonumber & \left.
+10 i r^3 K_{35}+3 r K_{42}+r^3 K_{43}-15 r K_{46}-10 r^3 K_{47}-r^5 K_{48}-i \Delta _2 r^2 K_{21}'+2 K_{22}'
\right. \\\nonumber & \left.
+8 r^2 K_{23}'-i r^2 K_{31}'+10 i r^2 K_{34}'+2 i r^4 K_{35}'+r K_{22}''+r^3 K_{23}''\right)\Big],
\end{align}
\begin{align}
V_h=&
\frac{3 g_f+2 g_f \vtu\cdot\vtd}{r^3}\Big[B_{\Lambda } r^2 \left(r J_{23}-J_{22}'\right) \vsd\cdot(\vr\times\vp)+i B_{\Lambda } r^2 \left(r J_{23}\right. 
 \\\nonumber & 
\left. -J_{22}'\right) (\vsu\cdot\vr) (\vsd\cdot\vp)+2 A_{\Lambda } M_N r^2 \left(r J_{23}-J_{22}'\right) (\vsu\times\vsd)\cdot\vr
 \\\nonumber & 
+B_{\Lambda } r^2 \left(-4 r J_{23}-5 i r J_{34}-i r^3 J_{35}-J_{22}'-r^2 J_{23}'\right) \vsu\cdot(\vr\times\vp)
 \\\nonumber & 
+B_{\Lambda } r^2 \left(4 i r J_{23}-5 r J_{34}-r^3 J_{35}+i J_{22}'+i r^2 J_{23}'\right) \vr\cdot\vp
 \\\nonumber & 
+2 A_{\Lambda } M_N r^2 \left(4 i r J_{23}-5 r J_{34}-r^3 J_{35}+i J_{22}'+i r^2 J_{23}'\right) (\vsu\cdot\vr)
 \\\nonumber & 
+\left(B_{\Lambda } r^2 \left(r \left(r J_{23}'-J_{22}''\right)+2 r J_{23}-J_{22}'\right)-i B_{\Lambda } r^2 \left(r J_{23}-J_{22}'\right) \vr\cdot\vp\right) \vsu\cdot\vsd
 \\\nonumber &
-B_{\Lambda } \left(r \left(r J_{23}'-J_{22}''\right)+J_{22}'\right) (\vsu\cdot\vr)(\vsd\cdot\vr)
-B_{\Lambda } r^2 \left(-3 i \Delta _1 r J_{21}+6 r J_{23}+3 \Delta _1 r J_{32}
\right. \\\nonumber & \left.
+\Delta _1 r^3 J_{33}+15 i r J_{34}+5 i r^3 J_{35}-3 r J_{42}-r^3 J_{43}+15 r J_{46}+10 r^3 J_{47}+r^5 J_{48}+2 J_{20}'
\right. \\\nonumber & \left.
-i \Delta _1 r^2 J_{21}'-4 J_{22}'
+2 r^2 J_{23}'+5 i r^2 J_{34}'+i r^4 J_{35}'+r J_{20}''-2 r J_{22}''\right)\Big],
\end{align}
\begin{align}
V_i=&
\frac{g_g}{r^3}\Big[2 A_{\text{$\Sigma $i}} M_N r^2 \left(-r J_{23}+J_{22}'\right) (\vsu\times\vsd)\cdot\vr
 \\\nonumber & 
+2 A_{\text{$\Sigma $i}} M_N r^2 \left(-6 i r J_{23}+5 r J_{34}+r^3 J_{35}-4 J_{11}'-4 i J_{22}'-2 i r^2 J_{23}'-r J_{11}''\right) (\vsu\cdot\vr)
 \\\nonumber & 
+B_{\text{$\Sigma $i}} r^2 \left(r \left(r J_{23}'-J_{22}''\right)+2 r J_{23}-J_{22}'\right) \vsu\cdot\vsd
\\\nonumber&
-B_{\text{$\Sigma $i}} \left(r \left(r J_{23}'-J_{22}''\right)+J_{22}'\right) (\vsu\cdot\vr)(\vsd\cdot\vr)
\\\nonumber & 
+B_{\text{$\Sigma $i}} r^2 \left(3 i \Delta _1^2 r J_{11}+3 i \Delta _1 \Delta _2 r J_{11}+6 i \Delta _1 r J_{21}+3 i \Delta _2 r J_{21}-3 \Delta _1^2 r J_{22}-3 \Delta _1 \Delta _2 r J_{22}
\right. \\\nonumber & \left.
-r \left(12+\Delta _1^2 r^2+\Delta _1 \Delta _2 r^2\right) J_{23}+3 i r J_{31}-6 \Delta _1 r J_{32}-3 \Delta _2 r J_{32}-2 \Delta _1 r^3 J_{33}-\Delta _2 r^3 J_{33}
\right. \\\nonumber & \left.
-30 i r J_{34}-10 i r^3 J_{35}-3 r J_{42}-r^3 J_{43}+15 r J_{46}+10 r^3 J_{47}+r^5 J_{48}+i \Delta _1^2 r^2 J_{11}'
\right. \\\nonumber & \left.
+i \Delta _1 \Delta _2 r^2 J_{11}'+2 i \Delta _1 r^2 J_{21}'+i \Delta _2 r^2 J_{21}'-2 J_{22}'-8 r^2 J_{23}'+i r^2 J_{31}'-10 i r^2 J_{34}'
\right.\\\nonumber& \left.
-2 i r^4 J_{35}'-r J_{22}''-r^3 J_{23}''\right)\Big].
\end{align}


We now take the previous expressions and reorganize 
the $\Lambda N\to NN$ potential into
the twenty spin and isospin operators of Eq.~(\ref{eq:nloops}).
The integrals coming from the different diagrams
depend on five different sets of $q_0$ and $q_0'$,
as listed on Table~\ref{tab:5q0q0p}.
The corresponding integrals are labeled with a superindex
$1,\dots,5$ according to the values they take from Table~\ref{tab:5q0q0p}.
\begin{table}[tb]
\centering
\begin{tabular}{cccccc}
       & 1 & 2 & 3 & 4 & 5
\\\hline
$q_0$    & $\Delta_1$  & $-\Delta_1$  & $-\Delta_1$  & $\Delta_1$  & $\Delta_1$ 
\\\hline
$q_0'$   &  $0$        & $-2\Delta_1$ & $\Delta_2$   & $-\Delta_1$ & $\Delta_1+\Delta_2$
\end{tabular}
\caption{Five possible values of $q_0$ and $q_0'$ that appear in the nine loop diagrams.}
\label{tab:5q0q0p}
\end{table}
For each spin operator we denote the scalar and isospin-isospin
potentials $V$ and $W$:

\begin{align}
V_1=&
g_b \left(r I_{23}^{1}-i {I_{11}^{1}}'\right) r^2-3 B_{\Lambda } g_f \left(J_{48}^4 r^5+K_{48}^2 r^5+i {J_{35}^4}' r^4-2 i {K_{35}^2}' r^4+15 i J_{35}^4 r^3\right. 
 \\\nonumber & 
\left. -J_{43}^4 r^3-K_{43}^2 r^3+2 K_{23}^2 \Delta _1^2 r^3+J_{33}^4 \Delta _1 r^3-K_{33}^2 \Delta _1 r^3-{K_{23}^2}'' r^3-i \Delta _1 {J_{21}^4}' r^2\right. 
 \\\nonumber & 
\left. -3 {J_{23}^4}' r^2-2 i \Delta _1^2 {K_{11}^2}' r^2+i \Delta _1 {K_{21}^2}' r^2+2 {K_{23}^2}' r^2+i {K_{31}^2}' r^2-24 J_{23}^4 r\right. 
 \\\nonumber & 
\left. -3 J_{42}^4 r+3 K_{23}^2 r+3 i K_{31}^2 r-3 K_{42}^2 r-3 i J_{21}^4 \Delta _1 r+3 J_{32}^4 \Delta _1 r+3 i K_{21}^2 \Delta _1 r\right. 
 \\\nonumber & 
\left. -3 K_{32}^2 \Delta _1 r-2 i {J_{11}^4}'' r+{J_{20}^4}'' r-i {K_{11}^2}'' r-4 i {J_{11}^4}'+2 {J_{20}^4}'-2 i {K_{11}^2}'\right)
 \\\nonumber & 
+\left(2 B_{\text{$\Sigma $32}}-\sqrt{3} B_{\text{$\Sigma $12}}\right) g_g \left(J_{48}^5 r^5-K_{48}^3 r^5-2 i {J_{35}^5}' r^4+2 i {K_{35}^3}' r^4\right. 
 \\\nonumber & 
\left. -J_{43}^5 r^3+K_{43}^3 r^3-J_{23}^5 \Delta _1^2 r^3-2 J_{33}^5 \Delta _1 r^3-J_{33}^5 \Delta _2 r^3+K_{33}^3 \Delta _2 r^3-J_{23}^5 \Delta _1 \Delta _2 r^3\right. 
 \\\nonumber & 
\left. -{J_{23}^5}'' r^3+{K_{23}^3}'' r^3+i \Delta _1^2 {J_{11}^5}' r^2+i \Delta _1 \Delta _2 {J_{11}^5}' r^2+2 i \Delta _1 {J_{21}^5}' r^2\right. 
 \\\nonumber & 
\left. +i \Delta _2 {J_{21}^5}' r^2+2 {J_{23}^5}' r^2+i {J_{31}^5}' r^2-i \Delta _2 {K_{21}^3}' r^2-2 {K_{23}^3}' r^2-i {K_{31}^3}' r^2\right. 
 \\\nonumber & 
\left. +3 J_{23}^5 r+3 i J_{31}^5 r-3 J_{42}^5 r-3 K_{23}^3 r-3 i K_{31}^3 r+3 K_{42}^3 r+6 i J_{21}^5 \Delta _1 r-6 J_{32}^5 \Delta _1 r\right. 
 \\\nonumber & 
\left. +3 i J_{21}^5 \Delta _2 r-3 J_{32}^5 \Delta _2 r-3 i K_{21}^3 \Delta _2 r+3 K_{32}^3 \Delta _2 r-i {J_{11}^5}'' r+i {K_{11}^3}'' r-2 i {J_{11}^5}'+2 i {K_{11}^3}'\right),
\end{align}

\begin{align}
W_1=&
r g_a \left(-4 \Delta _1 B_{10}^2+4 B_{20}^2+B^2 \Delta _1^2\right)+r g_c \left(-6 i I_{21}^{1}+6 I_{32}^{1}+2 r^2 I_{33}^{1}+r^2 I_{23}^{1} \Delta _1\right. 
 \\\nonumber & 
\left. -i r \Delta _1 {I_{11}^{1}}'-2 i r {I_{21}^{1}}'\right)-\frac{2}{3} \left(\sqrt{3} B_{\text{$\Sigma $12}}+2 B_{\text{$\Sigma $32}}\right) g_g \left(6 i r \Delta _1 J_{21}^5 +3 i r \Delta _2 J_{21}^5\right. 
 \\\nonumber & 
\left.+3 r J_{23}^5+3 i r J_{31}^5-3 r J_{42}^5-r^3 J_{43}^5+r^5 J_{48}^5-r^3 J_{23}^5 \Delta _1^2-6 r J_{32}^5 \Delta _1-2 r^3 J_{33}^5 \Delta _1\right. 
 \\\nonumber & 
\left. -3 r J_{32}^5 \Delta _2-r^3 J_{33}^5 \Delta _2-r^3 J_{23}^5 \Delta _1 \Delta _2+i r^2 \Delta _1^2 {J_{11}^5}'+i r^2 \Delta _1 \Delta _2 {J_{11}^5}'-2 i {J_{11}^5}'\right. 
 \\\nonumber & 
\left. +2 i r^2 \Delta _1 {J_{21}^5}'+i r^2 \Delta _2 {J_{21}^5}'+2 r^2 {J_{23}^5}'+i r^2 {J_{31}^5}'-2 i r^4 {J_{35}^5}'-i r {J_{11}^5}''-r^3 {J_{23}^5}''\right) 
 \\\nonumber & 
 +B_{\Lambda } \left(-r g_d \left(2 I_z^2 \Delta _1^3\right. \right. 
\left. \left. -7 I_{10}^2 \Delta _1^2+7 I_{20}^2 \Delta _1-r^2 I_{23}^2 \Delta _1+i r {I_{11}^2}' \Delta _1-6 i I_{21}^2-2 I_{30}^2+6 I_{32}^2\right. \right. 
 \\\nonumber & 
\left. \left. +2 r^2 I_{33}^2-2 i r {I_{21}^2}'\right)-2 g_f \left(J_{48}^4 r^5-K_{48}^2 r^5+i {J_{35}^4}' r^4+2 i {K_{35}^2}' r^4+15 i J_{35}^4 r^3\right. \right. 
 \\\nonumber & 
\left. \left. -J_{43}^4 r^3+K_{43}^2 r^3-2 K_{23}^2 \Delta _1^2 r^3+J_{33}^4 \Delta _1 r^3+K_{33}^2 \Delta _1 r^3+{K_{23}^2}'' r^3-i \Delta _1 {J_{21}^4}' r^2\right. \right. 
 \\\nonumber & 
\left. \left. -3 {J_{23}^4}' r^2+2 i \Delta _1^2 {K_{11}^2}' r^2-i \Delta _1 {K_{21}^2}' r^2-2 {K_{23}^2}' r^2-i {K_{31}^2}' r^2-24 J_{23}^4 r\right. \right. 
 \\\nonumber & 
\left. \left. -3 J_{42}^4 r-3 K_{23}^2 r-3 i K_{31}^2 r+3 K_{42}^2 r-3 i J_{21}^4 \Delta _1 r+3 J_{32}^4 \Delta _1 r-3 i K_{21}^2 \Delta _1 r\right. \right. 
 \\\nonumber & 
\left. \left. +3 K_{32}^2 \Delta _1 r-2 i {J_{11}^4}'' r+{J_{20}^4}'' r+i {K_{11}^2}'' r-4 i {J_{11}^4}'+2 {J_{20}^4}'+2 i {K_{11}^2}'\right)\right)
 \\\nonumber & 
+\frac{2}{3} \left(\sqrt{3} B_{\text{$\Sigma $12}}+B_{\text{$\Sigma $32}}\right) \left(r g_e \left(-2 \Delta _1^2 I_{10}^3+\Delta _1 \Delta _2 I_{10}^3
 -6 i I_{21}^3-2 I_{30}^3\right. \right. 
 \\\nonumber & 
\left. \left.+6 I_{32}^3+2 r^2 I_{33}^3+5 I_{20}^3 \Delta _1-r^2 I_{23}^3 \Delta _1-2 I_{20}^3 \Delta _2+i r \Delta _1 {I_{11}^3}' -2 i r {I_{21}^3}'\right)
\right. \\\nonumber & \left.
+g_g \left(-K_{48}^3 r^5+2 i {K_{35}^3}' r^4+K_{43}^3 r^3+K_{33}^3 \Delta _2 r^3+{K_{23}^3}'' r^3\right. \right. 
 \\\nonumber & 
\left. \left. -i \Delta _2 {K_{21}^3}' r^2-2 {K_{23}^3}' r^2-i {K_{31}^3}' r^2-3 K_{23}^3 r-3 i K_{31}^3 r+3 K_{42}^3 r-3 i K_{21}^3 \Delta _2 r\right. \right. 
 \\\nonumber & 
\left. \left. +3 K_{32}^3 \Delta _2 r+i {K_{11}^3}'' r+2 i {K_{11}^3}'\right)\right),
\end{align}

\begin{align}
V_{\vsu\cdot\vsd}=&
3 B_{\Lambda } g_f \frac1r\left(r^2 {J_{23}^4}'-i r {J_{11}^4}''+2 J_{23}^4 r-i {J_{11}^4}'+r^2 {K_{23}^2}'-i r {K_{11}^2}''+2 K_{23}^2 r-i {K_{11}^2}'\right)
 \\\nonumber & 
+g_g \left(2 B_{\text{$\Sigma $32}}-\sqrt{3} B_{\text{$\Sigma $12}}\right)\frac1r
 \left(r^2 {J_{23}^5}'-i r {J_{11}^5}''+2 J_{23}^5 r-i {J_{11}^5}'+r^2 {K_{23}^3}'
\right. \\\nonumber & \left.
 -i r {K_{11}^3}''+2 K_{23}^3 r-i {K_{11}^3}'\right),
\end{align}

\begin{align}
W_{\vsu\cdot\vsd}=&
2 B_{\Lambda } g_f\frac1r \left(r^2 {J_{23}^4}'-i r {J_{11}^4}''+2 J_{23}^4 r-i {J_{11}^4}'-r^2 {K_{23}^2}'+i r {K_{11}^2}''-2 K_{23}^2 r+i {K_{11}^2}'\right)
 \\\nonumber & 
-2 g_g \left(\sqrt{3} B_{\text{$\Sigma $12}}+2 B_{\text{$\Sigma $32}}\right) 
\frac{1}{3r}
\left(r^2 {J_{23}^5}'-i r {J_{11}^5}''+2 J_{23}^5 r-i {J_{11}^5}'\right)
 \\\nonumber & 
+2 g_g \left(\sqrt{3} B_{\text{$\Sigma $12}}+B_{\text{$\Sigma $32}}\right)
\frac{1}{3r}
\left(r^2 {K_{23}^3}'-i r {K_{11}^3}''+2 K_{23}^3 r-i {K_{11}^3}'\right),
\end{align}


\begin{align}
V_{\vsu\cdot\hr}=&
 -6 A_{\Lambda }M_N g_f \left(J_{35}^4 r^3-i r^2 {J_{23}^4}'+i J_{23}^4 r+{J_{11}^4}'+K_{35}^2 r^3-2 i r^2 {K_{23}^2}'-r {K_{11}^2}''-i K_{23}^2 r\right)
\nonumber \\ &
 2 M_N g_g \left(2 A_{\text{$\Sigma $32}}-\sqrt{3} A_{\text{$\Sigma $12}}\right) \left(J_{35}^5 r^3-2 i r^2 {J_{23}^5}'-r {J_{11}^5}''-i J_{23}^5 r+K_{35}^3 r^3-i r^2 {K_{23}^3}'\right. 
 \\\nonumber & 
\left.+i K_{23}^3 r+{K_{11}^3}'\right),
\end{align}

\begin{align}
W_{\vsu\cdot\hr}=&
 -2 g_d M_N A_{\Lambda }
 \left(\Delta _1 I_{11}^2 (-r)+2 I_{21}^2 r+i \Delta _1 {I_z^2}'-2 i {I_{10}^2}'\right)
 \\\nonumber & 
 -4 g_fM_N A_{\Lambda } \left(J_{35}^4 r^3-i r^2 {J_{23}^4}'+i J_{23}^4 r+{J_{11}^4}'-K_{35}^2 r^3+2 i r^2 {K_{23}^2}'+r {K_{11}^2}''+i K_{23}^2 r\right) 
 \\\nonumber & 
 +\frac{2}{3} \left(\sqrt{3} A_{\text{$\Sigma $12}}+A_{\text{$\Sigma $32}}\right) 
  \left[-2 r g_e M_N \left(2 I_{21}^3-\Delta _1 I_{11}^3\right)
\right.\\\nonumber & \left. 
  +2 g_gM_N \left(K_{35}^3 r^3-i r^2 {K_{23}^3}'+i K_{23}^3 r+{K_{11}^3}'\right)\right]
 \\\nonumber & 
 -\frac{4}{3} g_g r M_N\left(\sqrt{3} A_{\text{$\Sigma $12}}+2 A_{\text{$\Sigma $32}}\right) \left(J_{35}^5 r^2-2 i r {J_{23}^5}'-{J_{11}^5}''-i J_{23}^5\right),
\end{align}

\begin{align}
V_{\hr\cdot\hp}=&
-3 B_{\Lambda } g_f \left(J_{35}^4 r^3-i r^2 {J_{23}^4}'+i J_{23}^4 r+{J_{11}^4}'+K_{35}^2 r^3-2 i r^2 {K_{23}^2}'-r {K_{11}^2}''-i K_{23}^2 r\right),
\end{align}

\begin{align}
W_{\hr\cdot\hp}=&
 B_{\Lambda } g_d \left(\Delta _1 I_{11}^2 r-2 I_{21}^2 r-i \Delta _1 {I_z^2}'+2 i {I_{10}^2}'\right)
 \\\nonumber & 
 -2 B_{\Lambda }g_f \left(J_{35}^4 r^3-i r^2 {J_{23}^4}'+i J_{23}^4 r+{J_{11}^4}'-K_{35}^2 r^3+2 i r^2 {K_{23}^2}'+r {K_{11}^2}''+i K_{23}^2 r\right),
\end{align}

\begin{align}
V_{(\vsu\cdot\vsd)(\hr\cdot\hp)}=&
-3 i B_{\Lambda } g_f \left(J_{23}^4 r-i {J_{11}^4}'+K_{23}^2 r-i {K_{11}^2}'\right),
\end{align}

\begin{align}
W_{(\vsu\cdot\vsd)(\hr\cdot\hp)}=&
2 B_{\Lambda } g_f \left(-i J_{23}^4 r-{J_{11}^4}'+i K_{23}^2 r+{K_{11}^2}'\right),
\end{align}

\begin{align}
V_{(\vsu\times\vsd)\cdot\hr}=&
 6 M_NA_{\Lambda } g_f \left(J_{23}^4 r-i {J_{11}^4}'+K_{23}^2 r-i {K_{11}^2}'\right)
 \\\nonumber & 
  -2 M_N g_g \left(2 A_{\text{$\Sigma $32}}-\sqrt{3} A_{\text{$\Sigma $12}}\right) \left(J_{23}^5 r-i {J_{11}^5}'+K_{23}^3 r-i {K_{11}^3}'\right),
\end{align}

\begin{align}
W_{(\vsu\times\vsd)\cdot\hr}=&
 4M_N A_{\Lambda } g_f \left(J_{23}^4 r-i {J_{11}^4}'-K_{23}^2 r+i {K_{11}^2}'\right)
  \\\nonumber & 
 +\frac{4}{3}M_N g_g \left(\sqrt{3} A_{\text{$\Sigma $12}}
 +2M_N A_{\text{$\Sigma $32}}\right) \left(J_{23}^5 r-i {J_{11}^5}'\right)
 \\\nonumber & 
 -\frac{4}{3} M_Ng_g \left(\sqrt{3} A_{\text{$\Sigma $12}}+A_{\text{$\Sigma $32}}\right) \left(K_{23}^3 r-i {K_{11}^3}'\right),
\end{align}

\begin{align}
V_{\vsu\cdot(\hr\times\hp)}=&
3 B_{\Lambda } g_f \left(-i J_{35}^4 r^3-r^2 {J_{23}^4}'+J_{23}^4 r-i {J_{11}^4}'-i K_{35}^2 r^3-2 r^2 {K_{23}^2}'+i r {K_{11}^2}''-K_{23}^2 r\right),
\end{align}

\begin{align}
W_{\vsu\cdot(\hr\times\hp)}=&
B_{\Lambda }g_d \left(i \Delta _1 I_{11}^2 r-2 i I_{21}^2 r+\Delta _1 {I_z^2}'-2 {I_{10}^2}'\right)
 +2 g_f B_{\Lambda }
 \left(-i J_{35}^4 r^3-r^2 {J_{23}^4}'+J_{23}^4 r
\nonumber \right.  \\ & \left.
 -i {J_{11}^4}'+i K_{35}^2 r^3+2 r^2 {K_{23}^2}'-i r {K_{11}^2}''+K_{23}^2 r\right),
\end{align}

\begin{align}
V_{\vsd\cdot(\hr\times\hp)}=&
3 B_{\Lambda } g_f \left(J_{23}^4 r-i {J_{11}^4}'+K_{23}^2 r-i {K_{11}^2}'\right),
\end{align}

\begin{align}
W_{\vsd\cdot(\hr\times\hp)}=&
2 B_{\Lambda } g_f \left(J_{23}^4 r-i {J_{11}^4}'-K_{23}^2 r+i {K_{11}^2}'\right),
\end{align}

\begin{align}
V_{(\vsu\cdot\hr)(\vsd\cdot\hp)}=&
3 B_{\Lambda } g_f \left(i J_{23}^4 r+{J_{11}^4}'+i K_{23}^2 r+{K_{11}^2}'\right),
\end{align}

\begin{align}
W_{(\vsu\cdot\hr)(\vsd\cdot\hp)}=&
2 B_{\Lambda } g_f \left(i J_{23}^4 r+{J_{11}^4}'-i K_{23}^2 r-{K_{11}^2}'\right),
\end{align}

\begin{align}
V_{\hat{S}_{12}}=&
-\frac{3}{r} B_{\Lambda } g_f \left(r^2 {J_{23}^4}'-i r {J_{11}^4}''+i {J_{11}^4}'+r^2 {K_{23}^2}'-i r {K_{11}^2}''+i {K_{11}^2}'\right)
 \\\nonumber & 
-\frac{1}{r}g_g \left(2 B_{\text{$\Sigma $32}}-\sqrt{3} B_{\text{$\Sigma $12}}\right) \left(r^2 {J_{23}^5}'-i r {J_{11}^5}''+i {J_{11}^5}'+r^2 {K_{23}^3}'-i r {K_{11}^3}''+i {K_{11}^3}'\right),
\end{align}

\begin{align}
W_{\hat{S}_{12}}=&
-\frac{2}{r} B_{\Lambda } g_f \left(r^2 {J_{23}^4}'-i r {J_{11}^4}''+i {J_{11}^4}'-r^2 {K_{23}^2}'+i r {K_{11}^2}''-i {K_{11}^2}'\right)
 \\\nonumber & 
+\frac{2}{3r} g_g \left(\sqrt{3} B_{\text{$\Sigma $12}}+2 B_{\text{$\Sigma $32}}\right) \left(r^2 {J_{23}^5}'-i r {J_{11}^5}''+i {J_{11}^5}'\right)
 \\\nonumber & 
-\frac{2}{3r} g_g \left(\sqrt{3} B_{\text{$\Sigma $12}}+B_{\text{$\Sigma $32}}\right) \left(r^2 {K_{23}^3}'-i r {K_{11}^3}''+i {K_{11}^3}'\right).
\end{align}


\end{document}